\documentclass[twocolumn,showpacs,preprintnumbers,superscriptaddress,amsmath,amssymb,nofootinbib]{revtex4}
\usepackage{graphicx,color}
\usepackage[section]{placeins}
\newcommand{\cd}{\makebox[0.08cm]{$\cdot$}}
\newcommand{\bg}[1]{\mbox{\boldmath $#1$}}
\newcommand{\sla}{\not\!}
\begin{document}
\vspace{0.5cm}
\title{Systematic renormalization scheme in light-front dynamics \\
with Fock space truncation}
\author{V.A.~Karmanov}
\affiliation {Lebedev Physical Institute, Leninsky Prospekt 53,
119991 Moscow, Russia}
\author{J.-F.~Mathiot}
\affiliation {Laboratoire de Physique
Corpusculaire, Universit\'e Blaise-Pascal, \\
CNRS/IN2P3, 24 avenue des Landais, F-63177 Aubi\`ere Cedex,
France}
\author{A.V.~Smirnov}
\affiliation {Lebedev Physical Institute, Leninsky Prospekt 53,
119991 Moscow, Russia}

\bibliographystyle{unsrt}

\begin{abstract}
Within the framework of the covariant formulation of light-front
dynamics, we develop a general non-perturbative renormalization
scheme based on the Fock decomposition of the state vector and its
truncation. The counterterms and bare parameters needed to
renormalize the theory depend on the Fock sectors.
We present a general strategy in order to calculate these quantities,
as well as state vectors of physical systems, in a truncated Fock
space. The explicit dependence of our formalism on the orientation
of the light front plane is essential in order to analyze the
structure of the counterterms. We apply our formalism to the
two-body (one fermion and one boson) truncation in the Yukawa
model and in QED, and to the three-body truncation in a scalar
model. In QED, we recover analytically, without any perturbative
expansion, the renormalization of the electric charge, according
to the requirements of the Ward identity.
\end{abstract}
\maketitle

\section{Introduction} \label{intro}
The relevance of a coherent relativistic description of few-body
systems is now well recognized  in particle as well as in nuclear
physics. Concerning particle physics, a relativistic formalism is
necessary for the understanding of the various components of the
nucleon or pion state vectors in terms of valence quarks, gluons,
and sea quarks, as revealed, for instance, in exclusive reactions
at very high momentum transfer. The need for a coherent
relativistic approach to few-body systems has also become clear in
nuclear physics in order to check the validity of the standard
description of the microscopic structure of nuclei in terms of
correlated pion exchanges between nucleons within the general
framework of chiral perturbation theory. In this case,
electromagnetic interactions play a central role in "seeing" pion
exchanges in nuclei.

In the non-relativistic limit (when the speed of light $c$ goes to
infinity) a system of particles is described by its wave function
defined at fixed moments of time or, in other words, on the plane
$t=\mbox{const}$, and its time evolution is governed by the
Schr\"{o}dinger equation, once the Hamiltonian of the system is
known. Relativistic description admits some freedom in choosing
the space-like hyper-surface on which the state vector is
defined~\cite{dirac}. A possible choice is to take, for this
purpose, the same plane $t=\mbox{const}$ (the so-called "instant"
form of dynamics). This is however not very well suited for
relativistic systems, since this plane is not invariant under
Lorentz boosts. It is much more preferable to use Light-Front
Dynamics (LFD) which is of particular interest among various
approaches applied so far to study relativistic systems. In the
standard version of LFD, the state vector is defined on the plane
$t+\frac{\displaystyle{z}}{\displaystyle{c}}= 0$~\cite{dirac},
invariant with respect to Lorentz boosts along the $z$ axis.

Advantages of using LFD to describe physical systems are
well known. The main  one concerns the structure of the vacuum.
Because of kinematical constraints, the plus-component $p^+\equiv
p^0+p^3$ (we take hereafter $c=1$) of the four-momentum $p$ of any
particle state, both real and virtual,
 is always positive or null. This implies that the
vacuum state coincides with the free vacuum, and all intermediate
states result from fluctuations of the physical system. One can
thus construct any physical system in terms of combinations of
free fields, i.e. the state vector is decomposed in a series of
Fock sectors with an increasing number of constituents. This
enables a systematic calculation of state vectors of physical
systems and their observables.

Note that the triviality of the vacuum in LFD, mentioned above, does
not prevent from non-perturbative zero-mode contributions (states with
$p^+=0$, sometimes called the "vacuum sector") to field operators,
when physical systems with spontaneous symmetry breaking are
considered~\cite{hksw}. An application to the $\phi^4$
model in $1+1$ dimension has been done in Ref.~\cite{gw}.

While the Fock decomposition is non-perturbative, it is only
meaningful if it converges rapidly. One way to look at this
convergence for a simple but nevertheless physically relevant
system is to investigate, within LFD, the Wick-Cutkosky model: a
system of two scalar particles of mass $m$ interacting by the
exchange of a massless scalar particle. Independently, the same
system can be considered within the four-dimensional Feynman
approach by solving the Bethe-Salpeter equation in the ladder
approximation which includes exchanges of an infinite number of
scalar bosons in the intermediate state. Comparing the results of
both calculations~\cite{kw}, we can see that the two- and
three-body components of the state vector represent as much as $90
\%$ of its norm, for $m=1$ GeV and a coupling constant of $2\pi$
which gives the maximal binding. Such a simple test shows that
even in the worst case (a large coupling constant and the exchange
of a boson of zero mass) the Fock decomposition is meaningful and
may converge rapidly. This however should be analyzed in more
realistic calculations.

The decomposition of the state vector of any physical system in
terms of Fock sectors on the Light Front (LF) enables a very
intuitive interpretation of the physical state, since  each Fock
sector is reminiscent of a non-relativistic many-body wave
function.

The standard version of LFD has however a serious drawback, since
the equation of the LF plane $t+z=0$ is not invariant under
spatial rotations. As we shall see later on, the breaking of the
rotational invariance has many important consequences as far as
the construction of bound states with definite angular momentum is
concerned, or in the calculation of electromagnetic amplitudes.

To avoid such an unpleasant feature of standard LFD, we shall use
below the Covariant formulation of LFD (CLFD)~\cite{k76,cdkm},
which provides a simple, practical, and very powerful tool in order to
describe physical systems as well as their
 electromagnetic amplitudes. In this
formulation, the state vector is defined on the plane
characterized by  the invariant equation $\omega \cd x=0$, where
$\omega$ is an arbitrary light-like four-vector with $\omega
^2=0$. The  standard LFD  on the plane $t+z=0$ is recovered by
considering the particular choice $\omega=(1,0,0,-1)$. The
covariance of our approach is caused by the invariance of the LF
plane equation $\omega \cd x=0$ under any Lorentz transformation
of both $\omega$ and $x$. This implies in particular that $\omega$
cannot be kept the  same in any reference frame,  as it takes
place in the standard formulation of LFD with $\omega=(1,0,0,-1)$.

There is of course equivalence, in principle, between the standard
and covariant forms of LFD. Within the same approximation (or for
exact calculations) CLFD reproduces the results of standard LFD as
a particular case. The physical observables should coincide in
both approaches, though their derivation in CLFD in most cases is
much simpler and more transparent. The relation between CLFD and
standard LFD reminds that between the Feynman graph technique and
old-fashioned perturbation theory.

CLFD has first been used to investigate the general structure of
few-body systems and their electromagnetic observables in the tree
approximation (see Ref.~\cite{cdkm} for a review). If one wants to
go beyond this phenomenological analysis, one has to be able to
calculate the state vector of a physical system from a given
Hamiltonian in a non-perturbative framework.

\begin{figure}[btph]
\begin{center}
\includegraphics[width=20pc]{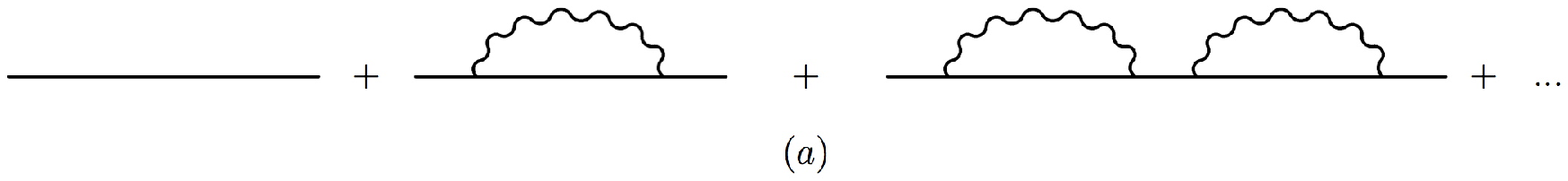}
\includegraphics[width=5pc]{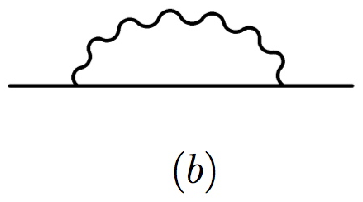}
\caption{Fermion propagator in the chain approximation, within the
two-body Fock space truncation (a) and an irreducible contribution
to the perturbative self-energy (b). Solid and wavy lines
correspond to fermions and bosons, respectively.\label{chain}}
\end{center}
\end{figure}

Consider, as an example, a system composed of interacting fermion
and bosons. In the simple two-body Fock space truncation, the
physical fermion state vector is represented as a sum of two
sectors: the one single fermion state and the one fermion plus one
boson state. The fermion propagator is thus given, in the chain
approximation, by the contributions indicated in
Fig.~\ref{chain}(a).
It is non-perturbative in the sense that it involves contributions
to all orders in the coupling constant $g$, but approximate, since
it incorporates at most two particles in the intermediate states.
It is well known that this infinite series can be summed up in
terms of the (perturbative) self-energy $\Sigma(p)$ of order
$g^2$, as indicated in Fig.~\ref{chain}(b). In this two-body
truncation, the equivalence between the LF fermion propagator
(calculated in CLFD) and the two-point Green's function
(calculated in the Feynman four-dimensional approach) has been
shown to occur very naturally to all orders in $g$~\cite{kms_04}.

The fermion propagator enters into the expression for the
observable fermion-boson scattering amplitude. This amplitude must
have a pole, in the $s$-channel, at $s=m^2$. To ensure such a
property, a Mass Counterterm (MC) must be added to the
self-energy. Besides that, the coupling constant coming into the
vertices of the diagrams in Fig.~\ref{chain} can not be identified
{\em a priori} with the physically observed quantity, but should
be treated as some bare (non-renormalized) parameter. In order to
calculate physical observables, the boson-fermion Bare Coupling
Constant (BCC) as well as the MC should be expressed in terms of
the physical coupling constant and the particle masses. This has
been done, for the two-body Fock state truncation in CLFD, in Ref.~\cite{kms_04}. However, a general
renormalization scheme needed to determine the MC and the BCC for
the most general case of Fock space truncation has not been
proposed yet.

Already at the level of  the two-body Fock space truncation, one
has to deal with loop diagrams [like the self-energy contribution
shown in Fig.~\ref{chain} (b)]. Their amplitudes diverge for high
internal momenta. The implementation of any renormalization scheme
essentially depends on the way of regularization of divergent
amplitudes. This is indeed a non-trivial task, as it has been
already mentioned in various contexts~\cite{kms_07, bj}. The
regularization of amplitudes in LFD by traditional cutoffs imposed
on the transverse and longitudinal components of particle momenta,
for instance, corresponds to restricting the integration volume by
a rotationally non-invariant domain. The regularized amplitudes
depend therefore not only on the size of this domain (i.e., on the
cutoff values), but also on its orientation determined by the
orientation of the LF plane.

Another source of violation of rotational invariance is the
Fock space truncation itself. As a consequence, the number and the
structure of the counterterms needed to renormalize the theory
depend on the LF plane orientation as well. CLFD allows us to
parameterize the latter dependence in a very transparent form,
through the four-vector $\omega$. Moreover, the covariant
formulation of the approach is mandatory in order to define what
are the physical parameters of the theory (and hence to be able to
renormalize the latter), since it enables an explicit separation
of any spurious contributions depending on $\omega$. This is the
case, for instance, for the two-body wave function, as we shall
see in Sec.~\ref{clfd}.

Following the analysis of Ref.~\cite{kms_07}, we choose the
Pauli-Villars (PV) regularization scheme in order to impart
mathematical sense to divergent amplitudes. This scheme also
preserves rotational invariance, as well as other important
symmetries like gauge invariance. Though the PV regularization
was developed initially for the four-dimensional Feynman approach,
it can be easily implemented into the LFD calculating machinery by
simply introducing additional fictitious PV fields \cite{bhm_01}.

The renormalization procedure must ensure that physical results
do not depend on the regularization parameters. Besides that, it
should be, first, non-perturbative and,
second, consistent with the truncation of the Fock
decomposition in the sense that it should not leave
any divergences uncancelled.

\begin{figure}[btph]
\begin{center}
\includegraphics[width=20pc]{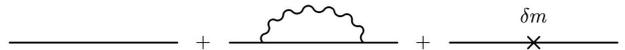}
\caption{Renormalization of the fermion propagator in the second
order of perturbation theory.\label{self}}
\end{center}
\end{figure}

Let us look, for example, at the renormalization of the fermion
propagator in the second order of perturbation theory. There exist
three contributions to the physical fermion propagator, as
indicated in Fig.~\ref{self}. These are, from left to right, the
free propagator, the self-energy contribution $\Sigma(p)$, and the
contribution from the MC $\delta m$. The sum of these three items
should be equal, at $p^2=m^2$, to the free propagator. This fixes
$\delta m = -\Sigma(p)$ at ${\sla p}=m$.

As we can see from Fig.~\ref{self}, Fock sectors with different
number of constituents are intimately connected through the
renormalization condition: the contribution of the MC (the last
diagram in Fig.~\ref{self}) corresponds to the one-body Fock
sector (a single fermion). It should however be opposite, at $\sla
p=m$, to the two-body (one fermion plus one boson) Fock sector
contribution given by the second diagram in Fig.~\ref{self}, in
order to cancel its divergence. This means that any MC or, more
generally, any bare parameter, should be associated with the
number of particles in a given  Fock sector. In other words, all
MC's and bare parameters must depend on the Fock sector under
consideration. This is a necessary condition.

Several attempts have already been made to address the problem  of
non-perturbative renormalization in the standard formulation of
LFD, either in the Yukawa model (a fermion coupled to scalar
bosons) or in QED, using various regularization schemes. Early
calculations were performed with a momentum cut-off for the Yukawa
model~\cite{ghp} and for QED~\cite{bhm_99}. As shown in
Ref.~\cite{kms_07}, the use of such a cut-off implies to consider
specific counterterms depending on the LF plane orientation.
Moreover, the absence of Fock sector dependent counterterms and
BCC's prevents any calculation to converge properly.

The use of PV fields to regulate the amplitudes has first been
advocated in Refs.~\cite{bhm_01} (with three PV bosons)
and~\cite{bhm_04} (with three PV fermions), for the Yukawa model
and QED, respectively. These calculations suffer however from the
lack of a non-perturbative procedure to determine the parameters
of the PV fields, as well as from an
incorrect chiral limit. Again, no Fock sector
dependent counterterms were considered, which left
divergences uncancelled. In particular, this prevents the two-body
calculation of QED  to reproduce the well known radiative
correction to the anomalous magnetic moment of  the electron
(the Schwinger correction).  We shall see in
Sec.~\ref{QED} how it arises naturally in our scheme.

Most recent calculations in the Yukawa  model with the
two-~\cite{bhm_03} and three-body~\cite{bhm_06} Fock space
truncations used simultaneously a PV fermion and a PV boson to
regulate the amplitudes. This regularization procedure is adequate
to preserve rotational invariance, at least for the two-body
truncation, according to the analysis of Ref.~\cite{kms_07}.
However these two calculations are plagued with uncancelled
divergences.

The dependence of the counterterms on the Fock sectors has been
first suggested in Ref.~\cite{wp} in the context of a simple
calculation within the two-body Fock space truncation. This idea
has however never been formulated as a coherent renormalization
scheme.

The  main aim of the present article is to develop such
renormalization scheme. We propose a complete and coherent
strategy to determine the counterterms and the bare parameters in
LFD calculations with a Fock space truncation to any order. A
preliminary account of such a scheme was presented in
Ref.~\cite{mks}. We conjecture that this renormalization scheme is
also sufficient to avoid any uncancelled divergences in any order
of the Fock space truncation, provided appropriate counterterms
necessary to recover rotational invariance (if needed) are taken
into account. We shall demonstrate below that this is indeed the
case for some model and realistic physical systems, within the
two- and three-body Fock space truncations.

The plan of our paper is the  following. In Sec.~\ref{clfd}, we
recall the main features of the description of bound state systems
in CLFD, taking the Yukawa model as an example.  We expose in
Sec.~\ref{renor} our new systematic renormalization scheme in CLFD
calculations with Fock space truncation. Applications of this
scheme to particular physical systems --- to the Yukawa model and
QED --- within the two-body Fock space truncation
(Sec.~\ref{QED}), and to a purely scalar model for the three-body
truncation (Sec.~\ref{scalar}) are then considered. We present our
concluding remarks and outline possible perspectives in
Sec.~\ref{conc}. Some technical derivations are given in
Appendices.

\section{Description of
physical systems in the covariant formulation of light-front
dynamics} \label{clfd} In order to show how our renormalization
scheme  should be applied to the analysis of physical systems, we
shall consider in the following study the Yukawa model, i.e. a
physical fermion composed of a bare fermion coupled to scalar
bosons. This system is interesting from several points of view. It
is not as simple as a super-renormalizable purely scalar model,
while it has many similarities with QED in  the Feynman gauge, at
least for the case of the simple two-body truncation. It is thus
easy to extend our results, as shown in Sec.~\ref{QED}.

\subsection{The Yukawa model.
Construction of the light-front interaction
Hamiltonian}\label{ham} 
The Lagrangian describing a system of
interacting spin-1/2 fermion and scalar boson fields, taking into
account the mass renormalization of the fermion, is
\begin{equation}
\label{lagrfull} {\cal L}  = {\cal L}_{F}+{\cal L}_{B}+{\cal
L}_{FB},
 \end{equation}
where the three terms on the r.-h.s. are, respectively, the
fermion, boson, and interaction parts,
\begin{subequations}
\begin{eqnarray}
\label{lagFfree}
{\cal L}_{F} & = &{\displaystyle  i\bar{\Psi}\gamma^{\nu }
\partial_{\nu }\Psi-m\bar{\Psi}\Psi,}\\
\label{lagBfree}
{\cal L}_{B} & = &{\displaystyle  \frac{1}{2}\left[\partial_{\nu }\Phi
\partial^{\nu }\Phi-\mu^2\Phi^2\right],}\\
\label{lagint}
 {\cal L}_{FB}& = &{\displaystyle  g_{0}\bar{\Psi}\Psi\Phi+\delta m\bar{\Psi}\Psi.}
\end{eqnarray}
\end{subequations}
Here $\Psi=\Psi(x)$ and $\Phi=\Phi(x)$ are the Heisenberg fermion and
boson field operators, $g_{0}$ is the BCC, analogous to the bare charge $e_0$ in
QED, $m$ is the physical fermion mass, $\mu$ is the
physical boson mass, and $\delta m$ is the fermion
MC.

As already advocated in Ref.~\cite{kms_04}, it is more appropriate
and physically sounded to construct Fock sectors in terms of free
fields corresponding to particles with their physical masses. In
that case, one does not have to determine the fermion bare mass
$m_0$ but rather a MC $\delta m=m_0-m$,  as it is well
known~\cite{ps}. This choice of the renormalization procedure for
the fermion mass is the only way to keep the basis constructed
from free fields to be the same in all Fock sectors. In
our renormalization scheme the bare parameters like $m_0$ depend
on the Fock sector in which they appear. If one assigned the bare
mass $m_0$ to the free fermion field, the latter would be
different in different Fock sectors. Taking the free fermion field
with the physical mass $m$, on the contrary, fixes it once and for
all, while dependence of renormalization parameters on the Fock
sectors is carried over to the MC $\delta m$. Moreover, one may hope 
that the Fock state expansion may converge more
rapidly with the use of a  fixed physical mass as compared  to a 
(divergent) bare mass.

Working in LFD, we have to deal with Hamiltonians, rather
than Lagrangians. Moreover, since we use Fock expansions in terms
of free fields, the Hamiltonian must be also expressed through
them (i.~e. taken in Schr\"{o}dinger or interaction
representation). The general procedure of deriving CLFD
Hamiltonians from Lagrangians is exposed in Ref.~\cite{kms_04}.
First, one should construct the energy-momentum tensor
\begin{equation}
\label{emt} \Theta_{\nu\rho}=\sum_i\left(\frac{\partial {\cal
L}}{\partial^{\nu}Y_i}\right)\partial_{\rho}Y_i-g_{\nu\rho}{\cal
L},
\end{equation}
where $Y_i$ denotes either $\Psi$ or $\bar{\Psi}$, or $\Phi$, the
sum running over all the fields, and the LF four-momentum
operator
\begin{equation}
\label{FMO} \hat{P}_{\rho}=\frac{1}{2}\int
d\sigma^{\nu}(x)\,\Theta_{\nu\rho},
\end{equation}
where the integration is performed on the three-dimensional space
element orthogonal to the "time" direction  (the role of time is
played in CLFD by the invariant combination $\omega\cd x$). The
four-momentum operator should then be expressed through the free
fields, taking into account constraints imposed on the field
components by the equations of motion. The corresponding operator
$\hat{P}_{\rho}$ can be represented as the sum
\begin{equation}
\label{pshr}
\hat{P}_{\rho}=\hat{P}^{(0)}_{\rho}+\hat{P}^{int}_{\rho},
\end{equation}
where the two terms on the r.-h.s. are, respectively, the free
(i.e. independent of the coupling constant and counterterms) and
interaction parts of the four-momentum. The operator
$\hat{P}^{int}_{\rho}$ is related to the interaction Hamiltonian
$H^{int}(x)$ by
\begin{equation}
\label{pintham} \hat{P}^{int}_{\rho}=\omega_{\rho}\int
H^{int}(x)\,\delta(\omega\cd x)\,d^4x.
\end{equation}

The calculations performed in Ref.~\cite{kms_04} showed that the
interaction Hamiltonian for the Yukawa model  includes also a set of so-called contact (or
instantaneous) terms which explicitly depend on the LF plane
orientation and essentially complicate calculations, both
perturbative and non-perturbative.

We shall use hereafter the PV regularization which not only
maintains rotational invariance, but also kills the contact terms,
as will be demonstrated below. The PV scheme can be easily
implemented into the Lagrangian~\cite{bhm_01} by introducing
additional fields (we will call them PV fields or  PV particles),
having negative norm, so that each physical field has its PV
counterpart. On the level of free Lagrangians, the physical and PV
fields are independent from each other, while they are mixed by
the interaction. The PV fermion and PV boson parts of the full
Lagrangian are
\begin{subequations}\label{lagFPVfree}
\begin{eqnarray}
{\cal L}_{F,PV} & = &{\displaystyle -
i\bar{\Psi}_{PV}\gamma^{\nu}\partial_{\nu}\Psi_{PV}+
m_1\bar{\Psi}_{PV}\Psi_{PV},}\\
{\cal L}_{B,PV} & = &{\displaystyle
-\frac{1}{2}\left[\partial_{\nu}\Phi_{PV}\partial^{\nu}\Phi_{PV}-
\mu_1^2\Phi_{PV}^2\right],}
\end{eqnarray}
\end{subequations}
with $m_1$ and $\mu_1$ being the PV fermion and PV boson masses.
Note that the Lagrangians~(\ref{lagFPVfree}) differ by a minus
sign from the Lagrangians~(\ref{lagFfree}) and (\ref{lagBfree})
for the physical fields. The interaction Lagrangian involves all
types of fields and has the form
\begin{equation}
\label{lagPVint}
 {\cal L}_{FB,PV}=g_{0}\bar{\Psi}'\Psi'\Phi'+\delta m \bar{\Psi}'\Psi',
\end{equation}
where
\begin{equation}
\label{fieldsp}
\Psi'=\Psi+\Psi_{PV},\quad \Phi'=\Phi+\Phi_{PV}.
\end{equation}
The interaction is constructed in such a way that the physical and
PV fields come into Eq.~(\ref{lagPVint}) on equal grounds. This
feature ensures the cancellation of ultra-violet divergencies. The
full Lagrangian combining the physical and PV
contributions is thus
\begin{equation}
\label{lagrPVfull} {\cal L}_{PV}  = {\cal L}_{F}+{\cal
L}_{B}+{\cal L}_{F,PV}+ {\cal L}_{B,PV}+{\cal L}_{FB,PV}.
 \end{equation}
The Lagrangian~(\ref{lagrPVfull}) generates the interaction
Hamiltonian
\begin{equation}
\label{hamPV}
H^{int}_{PV}(x)=-g_{0}\bar{\psi'}\psi'\varphi'-\delta
m\bar{\psi'}\psi'
\end{equation}
with $\psi'=\psi+\psi_{PV}$ and $\varphi'=\varphi+\varphi_{PV}$.
The fields $\psi$ and $\psi_{PV}$ ($\varphi$ and $\varphi_{PV}$)
satisfy the free Dirac (Klein-Gordon) equations, with the
corresponding masses, in contrast to $\Psi$ and  $\Phi$ which satisfy the full Heisenberg equation.
 The main steps leading to Eq.~(\ref{hamPV})
are pointed out in Appendix~\ref{appA}.

The Hamiltonian~(\ref{hamPV}) has the traditional spin structure,
except for the fact that the "elementary" fields $\psi'$ and
$\varphi'$ are the sums of the physical and PV fields. In other
words, it does not contain any contact terms specific for LFD and
explicitly depending on the LF plane orientation. This is a great
merit of the PV regularization scheme.

The Lagrangian~(\ref{lagPVint}), as well as the
Hamiltonian~(\ref{hamPV}), depends on the MC $\delta
m$ and on the BCC $g_0$. For simplicity, we consider
here the case with only one coupling constant
to be determined, but our scheme is completely general and can be
easily extended to the case where many types of interaction occur.\\

Apart from the MC and the BCC
 entering the original Lagrangian, one may also need new
counterterms, at the level of the LF Hamiltonian, in order to
restore the symmetries broken by the Fock space
truncation~\cite{ghp} or by the regularization method~\cite{kms_07}. We
have already analyzed in Ref.~\cite{kms_04} the structure of such
counterterms in CLFD, using, as examples, the Yukawa model and QED
for the case of the two-body truncation and the standard LF
regularization by means of transversal and longitudinal cutoffs. 
Due to the explicit covariance of
CLFD, the general structure of such counterterms can be exhibited
in terms of the orientation, $\omega$, of the LF plane. The
simplest counterterm which one may consider is given by
\begin{equation} \label{ctw}
Z_\omega \bar \psi \frac{m\sla \omega} {i\omega\cd \partial} \
\psi,
\end{equation}
where $Z_\omega$ is a constant and $1/i(\omega\cd\partial)$ is the
operator $1/i\partial^+$, Eq.~(\ref{1/d}), written in covariant
notations. This counterterm has a structure similar to that of the
MC and appears, in all diagrams, as a factor $Z_{\omega}m{\sla
\omega}/(\omega\cd p)$ on each internal fermion line (here $p$ is
the four-momentum assigned to the line). Other counterterms with
more involved structure may appear if one increases the number of
Fock components, giving rise to many-body vertex corrections. The
general renormalization scheme we propose in this paper can easily
embrace all types of counterterms.

\subsection{Covariant formulation of light-front dynamics} \label{angular}

In CLFD, the state vector is defined on the LF plane of general
orientation  $\omega\cd x=\xi$, where $\omega$  is  an arbitrary
four-vector restricted by the condition  $\omega^2=0$, and $\xi$
is the LF "time". We shall take $\xi=0$, for convenience.

Let us recall here, for completeness, how the state vector of a
physical system is constructed. In order to avoid congesting
notations, we do not consider for the moment PV fields. These
fields influence only the explicit form of dynamical operators,
but not the general results discussed in this section. PV
fields can be easily incorporated later, when we shall study particular
physical systems.

We are interested in the state vector, $\phi_\omega^{J\sigma}(p)$,
of a bound system. It corresponds to definite values for the
mass $M$, the four-momentum $p$, and the total angular momentum
$J$ with projection $\sigma$ onto the $z$ axis in the rest frame,
i.e., the state vector forms a representation of the Poincar\'e
group. This means that it satisfies the following eigenstate
equations:
\begin{subequations}
\begin{eqnarray}\label{kt14}
\hat{P}_{\rho}\ \phi_\omega^{J\sigma}(p)&=&p_{\rho}\ \phi_\omega^{J\sigma}(p),\\
\label{kt15}
\hat{P}^{2}\ \phi_\omega^{J\sigma}(p)&=&M^2\ \phi_\omega^{J\sigma}(p),\\
\label{kt16}
\hat{S}^{2}\ \phi_\omega^{J\sigma}(p)&=&-M^2\ J(J+1)\ \phi_\omega^{J\sigma}(p),\\
\label{kt17} \hat{S}_{3}\ \phi_\omega^{J\sigma}(p)&=&M\
\sigma\phi_\omega^{J\sigma}(p),
\end{eqnarray}
\end{subequations}
where $\hat{S}_{\rho}$ is the Pauli-Lubanski vector
\begin{equation}\label{kt18}
\hat{S}_{\rho}= \frac{1}{2}\epsilon_{\rho\nu\alpha\beta} \
\hat{P}^{\nu}\ \hat{J}^{\alpha\beta},
\end{equation}
and $\hat{J}$ is the four-dimensional angular momentum
operator which is represented, similarly to $\hat{P}_{\rho}$,
Eq.~(\ref{pshr}), as a sum of the free and interaction parts:
\begin{equation}
\label{kt2} \hat{J}_{\rho\nu}=\hat{J}^{(0)}_{\rho\nu}
+\hat{J}^{int}_{\rho\nu}.
\end{equation}
In terms of the interaction Hamiltonian, we have
\begin{equation}\label{kt5}
 \hat{J}^{int}_{\rho\nu}=\int H^{int}(x)(x_{\rho}\omega_{\nu} -x_{\nu}
\omega_{\rho}) \delta(\omega\cd x)\ d^4x.
\end{equation}

From the general transformation properties of both the  state
vector and the LF plane, it follows~\cite{k82} that
\begin{equation}\label{kt12}
\hat{J}^{int}_{\rho\nu} \ \phi_\omega^{J\sigma}(p)=
\hat{L}_{\rho\nu}(\omega)\phi_\omega^{J\sigma}(p),
\end{equation}
where
\begin{equation}\label{kt13}
\hat{L}_{\rho\nu}(\omega) =i\left(\omega_{\rho}
\frac{\partial}{\partial\omega^{\nu}} -\omega_{\nu}
\frac{\partial}{\partial\omega^{\rho}}\right).
\end{equation}
The equation~(\ref{kt12}) is called the {\it angular condition}.
We can now use it in order to replace the operator
$\hat{J}^{int}_{\rho\nu}$ entering into Eq.~(\ref{kt18}) by
$\hat{L}_{\rho\nu}(\omega)$. Introducing the notations
\begin{subequations}
\begin{eqnarray}\label{kt19}
\hat{M}_{\rho\nu} &=&\hat{J}^{(0)}_{\rho\nu} +\hat{L}_{\rho\nu}(\omega),\\
\label{kt20} \hat{W}_{\rho}&=&
\frac{1}{2}\epsilon_{\rho\nu\alpha\beta} \ \hat{P}^{\nu}\
\hat{M}^{\alpha\beta},
\end{eqnarray}
\end{subequations}
we obtain, instead of Eqs.~(\ref{kt16}) and~(\ref{kt17}):
\begin{subequations}
\begin{eqnarray}\label{kt21}
\hat{W}^{2}\phi_\omega^{J\sigma}(p)&=&-M^2J(J+1)\ \phi_\omega^{J\sigma}(p),\\
\label{kt22} \hat{W}_{3}\ \phi_\omega^{J\sigma}(p)&=&M\ \sigma\
\phi_\omega^{J\sigma}(p).
\end{eqnarray}
\end{subequations}
{\it These equations do not contain the interaction Hamiltonian,
once} $\phi$ {\it satisfies Eqs.}~(\ref{kt14}) {\it
and}~(\ref{kt15}). The construction of the wave functions of
states with definite total angular momentum becomes therefore {\it
a purely kinematical problem}. Indeed, the transformation
properties of the state vector under rotations of the coordinate
system is fully determined by its total angular momentum, while
the dynamical part of the latter is separated out by means of the
angular condition. The dynamical dependence of the wave functions
on the LF plane orientation now turns into their explicit
dependence on the four-vector $\omega$~\cite{cdkm}. Such a
separation, in a covariant way, of kinematical and dynamical
transformations is a definite advantage of CLFD as compared to
standard LFD on the plane $t+z=0$.

%
\subsection{General Fock decomposition of the state vector}
\label{gdsv} According to the general properties of LFD, mentioned
in the Introduction, we decompose the state vector of a physical
system in Fock sectors. Schematically, we have
\begin{equation}
\phi^{J\sigma }_{\omega}(p) \equiv \vert 1 \rangle +
\vert 2 \rangle +
 \dots + \vert n \rangle + \dots
\end{equation}
Each term on the r.-h.s. denotes a state with a fixed number of
particles from which the physical system can be constructed.  In
the Yukawa model the analytical form of the Fock decomposition is
\begin{widetext}
\begin{eqnarray}
\phi^{J\sigma
}_{\omega}(p)&=&\sum_{n=1}^{\infty
}\frac{(2\pi)^{3/2}}{(n-1)!}\sum_{\sigma'} \int\phi_{n,\sigma\sigma'}(k_{1}\ldots
k_{n},p,\omega\tau_{n}) a^{\dag}_{\sigma'}({\bf
k}_{1})c^{\dag}({\bf k}_{2})\ldots c^{\dag}({\bf k}_{n})|0\rangle \nonumber \\
\label{eq21} &\times &\delta^{(4)}(k_{1}+\ldots
+k_{n}-p-\omega\tau_{n})2(\omega\cd p)d\tau_{n }\prod_{l
=1}^n \frac{d^3k_{l
}}{(2\pi)^{3/2}\sqrt{2\varepsilon_{k_{l }}}},
\label{twobody}
\end{eqnarray}
\end{widetext}
where $\phi_{n,\sigma\sigma'}(\ldots)$ is the $n$-body LF wave function (Fock component)
describing the state made of one free fermion and
$(n-1)$ free bosons, $a^{\dag}$ ($c^\dag$) are the free fermion
(boson) creation operators, $\varepsilon_{k_{l}}=\sqrt{{\bf
k}_{l}^{2}+m_{l}^{2}}$, and $m_{l}$ is the mass of the particle $l$
with the four-momentum $k_{l}$. The combinatorial factor $1/(n-1)!$
 is introduced in order to take into account the identity
of bosons.\footnote{Usually the factor $1/\sqrt{(n-1)!}$ is used,
instead of $1/(n-1)!$. Our choice however allows to remove
additional combinatorial factors in the equations for the Fock
components, which would arise in the former case.} The variables
$\tau_n$ describe how far off the energy shell the constituents
are. As explained in Appendix \ref{appB}, the momentum $\omega \tau_n$
can be identified with a fictitious particle, called spurion.
In practical calculations, the infinite sum over $n$ is
truncated by retaining terms with $n$ which does not exceed a
given number $N$, while those with $n>N$ are neglected.
Decompositions analogous to Eq.~(\ref{twobody}) can be easily
written for the QED case~\cite{kms_04} or for a purely scalar
system~\cite{bckm}.

The normalization  condition for the state vector  is given by
\begin{equation} \label{norma}
\phi^{\dag\, J\sigma'}_{\omega}(p')\phi^{J\sigma}_{\omega}(p)
= 2 \varepsilon_p\ \delta_{\sigma, \sigma '}
\delta^{(3)}(\bf{p} - \bf{p'}).
\end{equation}
Being rewritten through the CLFD wave functions, it has the form
\begin{equation}
\label{normwf1} \sum_{n=1}^{\infty}I_n=1,
\end{equation}
where
\begin{eqnarray}
I_n&=&\frac{(\omega\cd p)}{(2\pi)^{3(n-1)}(n-1)!}\nonumber \\
&&\times\int \left(\prod_{l=1}^{n}\frac{d^3k_l}{2\varepsilon_{k_l}}\right)d\tau_n
\delta^{(4)}\left(\sum_{l=1}^nk_l-p-\omega\tau_n\right)\nonumber \\
\label{normwf2}
&&\times\sum_{\sigma,\sigma'}\phi_{n,\sigma\sigma'}^{\dag}\phi_{n,\sigma\sigma'}^{\vphantom
\dag}
\end{eqnarray}
is the relative contribution of the $n$-body sector to the full
norm. For shortness, we omitted the arguments of the wave
functions. The factor $1/(n-1)!$ in Eq.~(\ref{normwf2}) appears as
a combined effect caused by the presence of the same factor in
Eq.~(\ref{twobody}) and by the contraction of the creation and
annihilation operators, when calculating the l.-h.s. of
Eq.~(\ref{norma}).

As follows from the discussion in Sec.~\ref{angular}, the spin
structure of the wave functions $\phi_{n,\sigma \sigma'}$ is very
simple, since it is purely kinematical, but it should incorporate
$\omega$-dependent components in order to fulfill the angular
condition~(\ref{kt12}). It is convenient to decompose each wave
function $\phi_{n,\sigma \sigma'}$ into invariant amplitudes
constructed from the particle four-momenta (including the
four-vector $\omega$!) and spin structures (matrices, bispinors,
etc.). In the Yukawa model, for instance, we have
\begin{subequations}
\label{Yuphi}
\begin{eqnarray}
\label{oneone}
\phi_{1,\sigma\sigma'}&= &\psi_1\  \bar{u}_{\sigma'}(k_1)u_{\sigma}(p),\\
\phi_{2,\sigma\sigma'}&=&\bar{u}_{\sigma'}(k_1) \left[ \psi_2  +
\psi'_2\ \frac{m \sla \omega }{\omega \cd p}\right]
u_{\sigma}(p), \label{onetwo}
\end{eqnarray}
\end{subequations}
since no other independent spin structures can be constructed.
Here $u$'s are bispinors, $\psi_1$, $\psi_2$, and $\psi_2'$ are
scalar functions determined by the dynamics. For a spin $1/2$
system coupled to scalar particles, the number of invariant
amplitudes for the two-body Fock component coincides with the
number of independent amplitudes of the reaction $\mbox{spin}\ 1/2
+ \mbox{scalar} \to \mbox{spin} \ 1/2 + \mbox{scalar}$, which is
$(2 \times 2)/2=2$, due to parity conservation.

Note that the formulas~(\ref{twobody}), (\ref{normwf2}),
 and~(\ref{Yuphi}) are written for the state vector which
contains physical particles only. The use of the PV
regularization, strictly speaking, changes them.
However, their generalization is
straightforward.
We do not give here the corresponding
general relations, but give their particular forms when we proceed to the consideration
of concrete physical systems.

\subsection{Eigenstate equation} \label{eigen}
The equations for the Fock components can be obtained from
Eq.~(\ref{kt15}) by substituting there the Fock
decomposition~(\ref{twobody}) of the state vector $\phi(p)$
(here and below we will omit, for shortness, all indices in the notation of the
state vector) and calculating the matrix elements
of the operator $\hat{P}^2$ in Fock space. With the
expressions~(\ref{pshr}) and~(\ref{pintham}), we can easily
get the eigenstate equation~\cite{bckm}:
\begin{equation}\label{eq1b}
2(\omega\cd p)\int \tilde{H}^{int}(\omega\tau)\frac{d\tau}{2\pi}
\phi(p)= -\left[\left(\hat{P}^{(0)}\right)^2-M^{2}\right]\phi(p),
\end{equation}
where $\tilde H^{int}$ is the interaction Hamiltonian in momentum
space:
\begin{equation}
\label{hamG} \tilde{H}^{int}(\omega\tau)=\int
H^{int}(x)e^{-i(\omega\cd x)\tau}d^4x.
\end{equation}
For the Yukawa model with the PV regularization,
$H^{int}(x)$ is given by Eq.~(\ref{hamPV}).

According to the decomposition~(\ref{twobody}), the
conservation law for the momenta in each Fock component has the
form
\begin{equation} \label{k1n}
k_1+k_2+\cdots +k_{n}=p+\omega\tau_n.
\end{equation}
Hence, the action of the operator
$\left(\hat{P}^{(0)}\right)^2-M^2$ on the state vector reduces
to the multiplication of each Fock component by the
factor $(\sum_{l=1}^n k_l)^2-M^{2}=2(\omega\cd p)\tau_n$.
It is therefore convenient to introduce the notation
\begin{equation}
{\cal G}(p)=2(\omega\cd p)\hat{\tau}\phi(p),
\end{equation}
where $\hat \tau $ is the operator which, acting on a given
component $\phi_{n,\sigma\sigma' }$ of $\phi(p)$, gives $\tau_n
\phi_{n,\sigma\sigma' }$. ${\cal G}(p)$ has the Fock decomposition
which is obtained from Eq.~(\ref{twobody}) by the replacement of
the wave functions $\phi_{n,\sigma\sigma' }$ by the vertex
functions $\Gamma_{n }$ (which we will also refer to as the Fock
components) defined by
\begin{equation}\label{eq22}
\bar{u}_{\sigma'}(k_{1})\Gamma_{n}u_{\sigma}(p)=(s_{n}-M^2)\phi_{n,\sigma\sigma'}
\end{equation}
and $s_{n}=(k_{1}+\ldots k_{n})^2$.  Since for each Fock component
$s_{n}-M^2=2(\omega\cd p)\tau_{n}$, we can cast the eigenstate
equation in the form
\begin{equation}\label{eq3}
{\cal G}(p) = \frac{1}{2\pi}\int
\left[-\tilde{H}^{int}(\omega\tau)\right]\frac{d\tau}{\tau} {\cal G}(p)\ .
\end{equation}
The physical bound state mass $M$ is found from the condition that
the eigenvalue is 1.  This equation is quite general and
equivalent to the eigenstate equation~(\ref{kt15}).
It is non-perturbative.

The normalization integrals~(\ref{normwf2}) rewritten through the
vertex functions are
\begin{eqnarray}
I_n&=&\frac{(\omega\cd p)}{(2\pi)^{3(n-1)}(n-1)!} \nonumber \\
&&\times\int
\left(\prod_{l=1}^{n}\frac{d^3k_l}{2\varepsilon_{k_l}}\right)d\tau_n
\delta^{(4)}\left(\sum_{l=1}^nk_l-p-\omega\tau_n \right)\nonumber \\
&&\times \frac{\mbox{Tr}\{({\sla p}+M)\bar{\Gamma}_n({\sla
k}_1+m)\Gamma_n\}}{(s_n-M^2)^2}
\label{intGamma}
\end{eqnarray}
with $\bar{\Gamma}_n=\gamma^0\Gamma_n^{\dag}\gamma^0$.

Since the Hamiltonian~(\ref{hamPV}) has the ordinary structure in
terms of the fields $\psi'$ and $\varphi'$ (i.e. it does not
include any contact terms), we may apply, for subsequent
calculations, the standard CLFD graph technique rules~\cite{cdkm}
with minor changes (see Appendix~\ref{appB}).

\section{Systematic Renormalization Scheme in CLFD } \label{renor}
In the usual renormalization scheme, the bare
parameters\footnote{The term "bare parameters" means here the
whole set of parameters entering into the interaction Hamiltonian,
e.g. the BCC, the fermion MC, etc.} are determined by fixing some
physical quantities like the particle masses and the physical
coupling constant. The physical parameters are thus expressed
through the bare ones. This identification implies in fact the
following two important consequences which are usually never
clarified in LFD calculations, but are at the heart of
our scheme.

{(i)} In order to express the physical parameters through the bare
ones, and vice versa, one should be able to calculate observables
or, in other words, physical amplitudes. In LFD, any physical
amplitude is represented as a sum of partial contributions, each
depending on the LF plane orientation.  Since an observable
quantity can not depend on the choice of the LF plane, this
spurious dependence must cancel in the whole sum. Such a situation
indeed takes place, for instance, in perturbation theory, provided
the regularization of divergencies in LFD amplitudes is done in a
rotationally invariant way~\cite{kms_07}. In non-perturbative LFD
calculations which are always approximate (say, due to the Fock
space truncation we just use here) the dependence on the LF plane
orientation may survive even in physical amplitudes. For this
reason, the identification of such amplitudes with observable
quantities becomes ambiguous and expressing the amplitudes through
the physical parameters turns into a non-trivial problem.

When working in standard LFD on the plane $t+z=0$, one may think
that even approximate LFD amplitudes do not depend on the LF plane orientation.
As a matter of fact, they do, but this dependence is simply hidden. It reflects itself by
the non-invariance of the corresponding amplitudes: the result of calculation
is affected by the choice of the reference frame.

As we have detailed in Sec.~\ref{clfd}, CLFD is a unique  tool to
control this dependence
in terms of the arbitrary light-like
four-vector $\omega$ specifying the LF plane orientation.
We shall see in Sec.~\ref{QED} how one should make use of this
property to define the physical fermion-boson coupling constant from the
two-body Fock component  of the type~(\ref{onetwo}).

{(ii)} The explicit form of the relationship between the bare and
physical parameters depends on the approximation which is made.
This is trivial in perturbation theory  where the order of
approximation is distinctly determined by the power of the
coupling constant. In our non-perturbative approach based on the
truncated Fock decomposition an analogous parameter is absent. At
the same time, to make calculations compatible with the order of
truncation, one has to trace somehow the level of approximation.
This implies that, on general grounds, the bare parameters should
depend on the Fock sector in which they are considered.  Moreover,
this dependence must be such that all  divergent contributions are
cancelled.

We will show in the following how to realize  the renormalization procedure in practice.
For clarity, we take, as a background, a model of interacting fermions and bosons like the
Yukawa model or QED.
At the same time, let us emphasize once more that the general
renormalization strategy developed here is applicable to physical systems with arbitrary
interaction admitting Fock decomposition of the state vector.

\subsection{Fock sector dependent counterterms}
\begin{figure}[btph]
\begin{center}
\includegraphics[width=18pc]{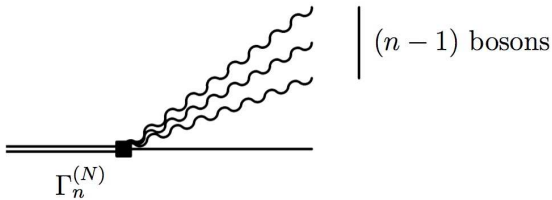}
\caption{Vertex function of order $n$ for the $N$-body Fock space
truncation. \label{gamman}}
\end{center}
\end{figure}

Let us call $N$ the maximal number of Fock sectors considered in a
given approximation, and $n$ the number of constituents in a given
Fock sector [one fermion and $(n-1)$ bosons]. We have $n\le N$.
Each Fock sector is described by its vertex function defined by
Eq.~(\ref{eq22}). In a truncated Fock decomposition, each vertex
function should depend on $N$. We shall thus denote it by
$\Gamma_n^{(N)}$. Graphically, $\Gamma_n^{(N)}$ can be represented
by the diagram shown in Fig.~\ref{gamman}.
%

\subsubsection{Mass counterterm}
The simple example of the fermion self-energy renormalization by
the MC within the two-body Fock space truncation, presented in the
Introduction, can serve as a guideline to define our general
rules. In this example, the MC should be labelled with a subscript
and denoted by $\delta m_{2}$, in order to indicate that it is
introduced in order to cancel, at $p\!\!\! /=m$, the fermion
self-energy contribution which belongs to the two-body Fock
sector.  In other words, $\delta m_2$ is related, by construction,
to the two-body state, even though it is attached to a single
fermion line. More generally, the subscript at $\delta m$
corresponds to the maximal number of particles in which the
fermion line where the MC is attached
 can fluctuate, so that the total number of particles at any LF time
equals $N$. In the given example  it is $2$.

Let thus denote by $\delta m_{l}$ the MC  in the
most general case.  Since we
truncate our Fock space to order $N$, one should make sure
that, at any LF time, the total number of particles is at
most $N$. Our first rule is thus:
\begin{itemize}
\item {\it in any amplitude where the MC  $\delta m_{l}$
appears, the value of $l$ is such that the total number of bosons
in flight plus $l$ equals the maximal number of Fock sectors
considered in the calculation, {\it i.e.} $N$.}
\end{itemize}

\begin{figure}[btph]
\begin{center}
\includegraphics[width=18pc]{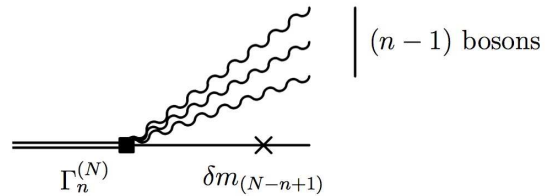}
\caption{Typical MC insertion. \label{gammadeltan}}
\end{center}
\end{figure}

For instance, in the typical contribution indicated in
Fig.~\ref{gammadeltan},  the MC  is
 $\delta m_{(N-n+1)}$. Indeed, since there are already $(n-1)$
bosons in flight, the fermion line can fluctuate in at most $l =
(N-n+1)$ particles, so that the total number of particles at a
given LF time is just $(N-n+1) + (n-1)=N$ and no more.

In order to calculate the whole set of the MC's $\delta m_{l}$
 one should proceed in the following way. Any
calculation of the state vector within the Fock space truncation of
order $N$ involves the MC's $\delta m_{l}$ with
$l=1,2,\dots N$. We emphasize at this point that the MC's
$\delta m_{l}$ are successive approximations to the true MC $\delta m$
appearing in the original Hamiltonian.  They are connected to each other
by some kind of recursion in the sense that finding $\delta m_l$
requires knowledge of the lower order counterterms,
i.e. $\delta m_1$, $\delta  m_2$, .. $\delta m_{l-1}$.
 This is analoguous to what happens in any
perturbative calculation where each MC relates to a definite order
of perturbation theory.

For the MC of lowest order, we simply have
\begin{equation}
\label{del12_1} \delta m_{1}=0,
\end{equation}
which is trivial since if the fermion can not fluctuate in more
than one particle, its mass is not renormalized at all. The
subsequent MC's are calculated by solving successively the
eigenvalue equations for the Fock components for the $2,3,\dots
N$-body Fock space truncations.
%

\subsubsection{Bare coupling constant}
\label{BCCg}
Let us now come to the determination of the BCC.  The general strategy
we developed above for the
calculation of the MC should of course be also
applied to the BCC case, with however a
bit of caution, since the BCC may enter in two
different types of contributions.

\begin{itemize}
\item The first one appears in the calculation of the state vector itself,
 when Eq.~(\ref{eq3}) is solved.
In that case, any boson-fermion coupling constant is associated
with the  emission or the absorption of a boson which participates
in the particle counting, in accordance with the rules detailed
above, since it is a part of the state vector.

\item The second one appears in the calculation of  the boson-fermion scattering
amplitude or of the boson-fermion three-point Green's function (3PGF)  like the
electromagnetic form factor in  the case of QED. This observable
is usually considered, at some kinematical point, to define the physical coupling
constant. Now the
external boson is an (asymptotic) free field rather than a part
of the state vector. The particle counting rule advocated above
should therefore not include the external boson line.
\end{itemize}

One has thus to distinguish two types of BCC's: $g_0$ and
$\bar{g}_0$ describing, respectively, the interaction vertices
of the fermion with internal and external bosons. As
we shall see below, these two BCC's are found from different
conditions and do not coincide with each other for a finite
order Fock space truncation.
\begin{figure}[btph]
\begin{center}
\includegraphics[width=7pc]{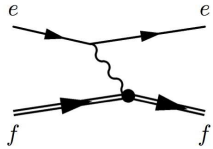}
\caption{Typical contribution to the two-body scattering amplitude
of a probe on a bound state system.\label{scat}}
\end{center}
\end{figure}
The necessity to introduce two BCC's can be also explained from
another point of view. Let us consider the scattering amplitude of
a given probe on a bound state system. Such scattering amplitude
can be represented by the diagram in Fig.~(\ref{scat}). The in and
out asymptotic states are defined in terms of the (structureless)
probe denoted by $e$ and the bound state system denoted by $f$.
The state vector of the bound system is calculated within a given
approximation (the Fock space truncation, in our case), starting
from a known Hamiltonian. Therefore, the calculated state vector
"knows" nothing about the subsequent interaction it can have with
the probe, i.e. it should be independent of any coupling to the
external virtual bosons exchanged between the probe and the bound
system.
\begin{figure}[btph]
\begin{center}
\includegraphics[width=18pc]{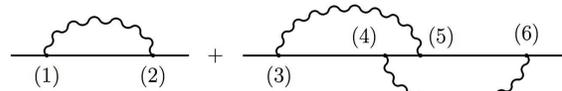}
\caption{Two- and three-body Fock sector contributions to the fermion self-energy.
 \label{radc}}
\end{center}
\end{figure}

 Similarly to the MC, the BCC's should also keep track
of the Fock sector in which they appear. To illustrate this fact,
let us write down some typical contributions to the fermion
self-energy, involving at most two bosons in flight (i.e. for
$N=3$).  They are shown in Fig.~\ref{radc}. All the vertices are
described by the internal BCC's, since the self-energy is a part
of the fermion state vector. The vertices $(1)$ and $(2)$ involve
the BCC's attached to the two-body sector in three-body truncated
Fock space. So, each of these vertices or both can be "dressed" by
one more bosonic loop, as indicated for the vertices $(4)$ and
$(5)$. The latter vertices correspond to states fully "saturated"
with bosons. In other words, no radiative corrections to them are
allowed in the given approximation. From here it follows that the
vertices $(1)$--$(3)$ and $(6)$, on the one hand, and $(4)$ and
$(5)$, on the other hand, are described by different BCC's.
Analogously to the MC, we will denote each internal BCC by
$g_{0l}$, where the subscript $l$ indicates which Fock sector the
given BCC belongs to. We can then formulate the general rule:
\begin{itemize}
\item {\it in any amplitude which couples constituents inside the state
vector one should attach to each vertex  the internal BCC
 $g_{0l}$. The value of $l$ is such
that the total number of bosons in flight before (after) the
vertex - if the latter corresponds to the boson emission
(absorbtion) - plus $l$ equals the maximal number of the Fock
sectors considered in the calculation, {\it i.e.} $N$.}
\end{itemize}
Applying this rule to the diagrams shown in Fig.~\ref{radc},  we
find that the vertices $(1)$--$(3)$ and $(6)$ are described by the
BCC $g_{03}$ [no bosons in flight before the vertices $(1)$ and
$(3)$ or after the vertices $(2)$ and $(6)$], while the vertices
$(4)$ and $(5)$ are described by the BCC $g_{02}$ (one boson in
flight before or after each vertex). The BCC $g_{02}$ is
calculated within the two-body Fock space truncation and enters
into the calculations within the three-body truncation as a known
parameter, while $g_{03}$ should be found. When considering the
four-body truncation, one should then calculate $g_{04}$, knowing
both BCC's $g_{02}$ and $g_{03}$. The procedure can thus be
extended to arbitrary $N$.
\begin{figure}[btph]
\begin{center}
\includegraphics[width=19pc]{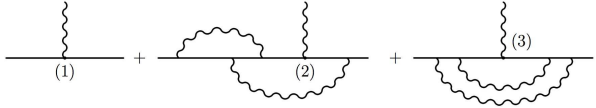}
\caption{Typical contribution to the fermion-boson 3PGF.
\label{elm2}}
\end{center}
\end{figure}

Let us finally consider typical contributions  to the 3PGF,
analogous to those for the self-energy, as discussed above. Some
of them are presented in Fig.~\ref{elm2}. The crucial difference
with the self-energy case consists in that the bosons emitted from
the vertices $(1)$--$(3)$ are absorbed by an external particle
(typically an external probe) which is not included into the state
vector. For this reason, the vertices $(1)$--$(3)$ are described
by the external BCC's $\bar{g}_{0l}$, while all the others
correspond to the internal BCC's $g_{0l}$. The values of $l$ for
the "internal" vertices are determined according to the rule
specified above, with the only exception that the external boson
does not take part in the particle counting. For the "external"
vertices $(1)$--$(3)$ the situation is quite similar. The vertices
$(1)$ and $(2)$ corresponding to the two-body Fock sector (the
external boson is not counted!) admit additional corrections,
within the three-body Fock space truncation, from "internal"
bosonic loops, while the vertex $(3)$ does not, since it appears
within the three-body sector. We can conclude from here that the
external BCC's attached to the vertices $(1)$ and $(2)$ should
differ from that for the vertex $(3)$.  We can thus formulate  the
following general rule:
\begin{itemize}
\item {\it in any amplitude which couples constituents of the
state vector with an external field, one should attach to the
vertex involving this external field the external BCC
$\bar{g}_{0l}$. The value of $l$ is such that, at the LF time
corresponding to the vertex, the total number of internal bosons
in flight - i.e. those emitted and absorbed by particles entering
the state vector - plus $l$ equals the maximal number of the Fock
sectors considered in the calculation, {\it i.e.} $N$.}
\end{itemize}
This rule prescribes to attach to the vertices $(1)$ and $(2)$ in
Fig.~\ref{elm2} the BCC $\bar{g}_{02}$, while the vertex $(3)$ is
associated with $\bar{g}_{01}$.

The lowest order BCC's are
\begin{subequations}
\begin{eqnarray}
g_{01}&=&0,\label{barg01}\\
\bar{g}_{01}&=&g.\label{g01}
\end{eqnarray}
\end{subequations}
Eq.~(\ref{barg01}) is trivial, because no fermion-boson
interaction is allowed in the one-body Fock space truncation.
Eq.~(\ref{g01}) reflects the fact that the external BCC, in the
same approximation, is not renormalized at all since a single
fermion can not be "dressed".

Note that the rule for attaching BCC's to external vertices holds
unchanged if the external field is a particle of another sort than
the bosons entering into the state vector. Such a situation takes
place, for instance, when one calculates the electromagnetic
vertex of a fermion, when the state vector does not contain
photons (e.g. in the Yukawa model). Evidently no problems with
particle counting arise in this case.

\begin{figure}[btph]
\begin{center}
\includegraphics[width=19pc]{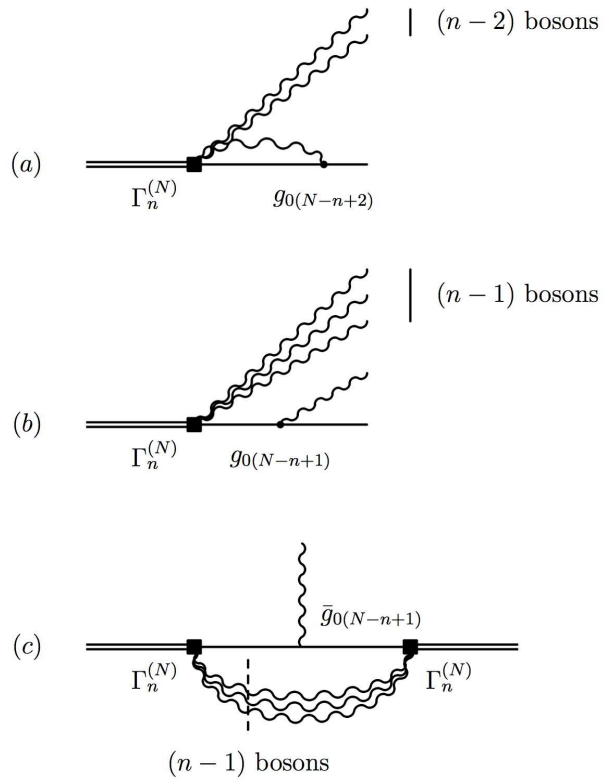}
\caption{Typical contributions to the fermion state vector for the
absorption (a) and the emission (b) of an internal boson, and to
the fermion-boson 3PGF (c). \label{gammag0n}}
\end{center}
\end{figure}

Some illustrations of the rules concerning the internal and
external BCC's are given in Fig.~\ref{gammag0n}.

This completes the set of our general rules to define in a
systematic and non-perturbative way the MC and the BCC's in LFD
calculations within truncated Fock space. The three rules exposed
above have very similar logical grounds based on the particle
counting in intermediate states. Namely, the index $l$ at the MC
and the BCC's is always calculated by the same rule: $l$ equals
the difference between the order of approximation $N$ and the
number of internal bosons in flight at the corresponding LF time.

Though we rely on the fermion-boson model when weconsidered 
the above procedure, this latter can be easily extended to other systems
with additional counterterms and bare parameters.

\subsection{Renormalization Conditions} \label{renorcond}

Once proper bare parameters and counterterms  have been
identified, one should fix them from a set of renormalization
conditions. In perturbation theory, there are three quantities to
be determined: the fermionic MC, the BCC, and the norm of the
fermion field (if the boson field renormalization is not
considered). Usually, the on-mass-shell renormalization is
applied, with the following conditions. The MC is fixed from the
requirement that the two-point Green's function has a pole at
$p^2=M^2$, where $M$ is the physical mass of the fermion. 
The
fermion field normalization is fixed by the condition that the
residue of the two-point Green's function in the pole is $1$. The
BCC is determined by requiring that the on-mass-shell 3PGF is
given by the product of the physical coupling constant and the
elementary (i.e. not "dressed") vertex.

One should now extend these conditions in order to determine the
bare parameters and the counterterms in a non-perturbative LFD
framework. We do not need to renormalize the fields in our
approach, since we deal with already normalized state vectors. It
remains to specify the procedure to find the set of MC's
$\delta m_{l}$ and the BCC's $\bar{g}_{0l}$ and $g_{0l}$.

The eigenstate equation for the state vector in the $N$-body
approximation includes two unknown parameters, the MC $\delta m_N$
and the BCC $g_{0N}$, while all $\delta m_l$ and $g_{0l}$ with
$l\leq N-1$ are  defined from calculations made for Fock space
truncations of lower orders. Provided $g_{0N}$ is fixed, the MC
$\delta m_N$ is found from the condition that the eigenvalue in
Eq.~(\ref{eq3}) equals 1 in the limit where the mass of the ground state, $M$, is equal to the mass of the fermionic constituent, $m$. To determine $g_{0N}$, one should relate
it to the physical coupling constant. For instance, in the
traditional QED renormalization scheme, the BCC is found from the
requirement that the residue of the photon-electron scattering
amplitude in the pole $s=m^2$, where $s$ is the invariant energy
squared of the system electron + photon, would be proportional to
the physical electron charge squared. This condition is not
restricted to perturbation theory and can be directly extended to
non-perturbative approaches as well. In our language, it is
equivalent to require that the two-body vertex $\Gamma_2$ is
proportional, at $s=m^2$, to the elementary vertex, the
proportionality coefficient being just the physical coupling
constant. For example, one has to require, at $s=m^2$, that
$\Gamma_2=g$ in the Yukawa model, and
$\Gamma_2^{\nu}=e\gamma^{\nu}$ in QED. From here a relation
between the internal BCC and the physical coupling constant can be
derived.

To determine the external BCC, a similar condition should be
imposed on the 3PGF. Indeed, the two-body vertex and the 3PGF
represent two different channels of the same reaction. Hence,
taken entirely on the mass and energy shells, they must coincide.
It follows from here that the 3PGF $G(p,p')$, where $p$ ($p'$) is
the four-momentum of the incoming (outgoing) fermion, reduces, at
$p^2={p'}^2=m^2$ and $q^2\equiv (p'-p)^2=\mu^2$, to the same
product of the elementary vertex and the physical coupling
constant. Since the internal BCC has already been found from the
calculation of $\Gamma_2$, the above condition on the 3PGF allows
to relate the external BCC with the physical coupling constant.

Note that the analytical continuation of the two-body vertex
function and the 3PGF to the non-physical points $s=m^2$ and
$q^2=\mu^2$ does not encounter any technical difficulties, even in
numerical calculations, since  we can use for this aim the
eigenstate equation~(\ref{eq3}). Its l.-h.s. contains the function
to be continued in the non-physical point. From the r.-h.s., both
$\Gamma_2$ and 3PGF are expressed through integrals involving the
vertex functions in the physical domain, whereas the dependence of the
integrand on $s$ and $q^2$ is explicit. We can put there
$s=m^2$ and $q^2=\mu^2$.

In perturbation theory, the equivalence of the on-mass-shell
two-body vertex function and 3PGF (calculated in the same order!)
appears automatically, as a consequence of the analytical
properties of scattering amplitudes. For this reason, it does not
make any sense to distinguish the external and internal BCC's,
because they are equal to each other to any order. The same would
happen in exact calculations, if they were possible. In our
non-perturbative approach based on truncated Fock decompositions, the non-renormalized
two-body vertex function and the 3PGF, even taken on the energy
shell, do not automatically coincide ,
in any finite order approximation. Moreover, they may be not
constant (i.e. keep dependence on particle momenta) and depend
also on the LF plane orientation (on the four-vector $\omega$, in
CLFD). If so, the question how to identify such objects with physical
constants in the renormalization point requires special
consideration.

First of all, one has to fix unambiguously all kinematical
variables in the renormalization point, in order that both the
two-body vertex function and the 3PGF would turn into constants.
Concerning the 3PGF, we can offer a universal solution of this
problem. Indeed, $G(p,p')$ depends dynamically on two scalar
variables, say, $q^2$ and $\omega\cd q/\omega\cd p$. The condition
$q^2=\mu^2$ fixes the first variable, but leaves free the second
one. However, if one imposes the condition $\omega\cd q=0$
(analogous to $q^+=0$ in ordinary LFD) on the four-vector
$\omega$, the dependence of the 3PGF on the second variable drops
out, and it becomes a constant at fixed $q^2$. This condition is
also necessary for the factorization of the total scattering
amplitude of a probe on the physical system under study, in terms of
the external boson propagator and the 3PGF. In practice, one should calculate
$G(p,p')$, keeping $\omega\cd p=\omega\cd p'$, and then continue
it analytically, as a function of only one variable $q^2$ to the
point $q^2=\mu^2$.

The two-body vertex $\Gamma_2$ also depends dynamically on two
scalar variables, e.g. $s$ and $x\equiv \omega\cd k_2/\omega\cd
p$, where $k_2$ is the boson four-momentum. The condition $s=m^2$
does not fix $x$. In an exact calculation, $\Gamma_2$ at $s=m^2$ does not depend on $x$, while the $x$-dependence survives because of approximations. Therefore, some additional restriction must be
imposed on $x$. Unfortunately, a universal choice how to
fix $x$ is hardly possible, in contrast to the case of the 3PGF.
This problem should be solved separately, for each particular
physical system.

Once the kinematics in the renormalization point is fixed, both
the two-body vertex function and the 3PGF turn, in this point,
into constant matrices.  Their dependence on the LF plane
orientation may however survive. For example, in the Yukawa model,
the term $\psi'_2\ {m \sla \omega }/{(\omega \cd p)}$ in the
two-body wave function~(\ref{onetwo}), which explicitly depends on
$\omega$, implies analogous dependence in the two-body vertex
function, unless $\psi'_2=0$ in the renormalization point. To get
rid of it, one should insert new counterterms into the LF
Hamiltonian, also explicitly depending on $\omega$, which cancel
completely the $\omega$-dependent term in $\Gamma_2$. Additional
counterterms may also be needed to kill possible
$\omega$-dependence of the 3PGF. According to the general
renormalization strategy, these counterterms must depend on the
Fock sector under consideration, in full analogy with the MC and
the BCC's.

Introducing the internal and external BCC's and imposing the above
renormalization conditions on both the on-energy-shell two-body
vertex and 3PGF, we force their coincidence, after
renormalization, for arbitrary Fock space truncation of finite
order. At this level, we restore the cross-invariance of
scattering amplitudes in the renormalization point.

Summarizing, we propose the following non-perturbative
renormalization conditions:

{\it
\begin{itemize}
\item The MC is fixed by solving the eigenstate equation
(\ref{eq3}) in the limit $M\to m$, where $M$ is the mass of the ground state of the physical system, and $m$ is the physical mass of the fermionic constituent. 
\item The state vector is
normalized according to the standard condition~(\ref{norma}).
\item The internal bare coupling constant $g_{0l}$ is fixed from
the condition that the $\omega$-independent part of the two-body
vertex function taken at $s=m^2$  and at a given value of $x$, denoted by $x^*$, is proportional to the elementary
vertex, with the proportionality coefficient being the physical
coupling constant. \item The external bare coupling constant
$\bar{g}_{0l}$ is fixed from the condition that the
$\omega$-independent part of the 3PGF calculated at $\omega\cd
q=0$ and taken at $q^2=\mu^2$ is proportional to the elementary
vertex, with the proportionality coefficient being the physical
coupling constant. \item The $\omega$-dependent counterterms in
the Hamiltonian are fixed by the conditions that the
$\omega$-dependent parts of the two-body vertex function in the point
($s=m^2,x^*)$ and the 3PGF in the point $(q^2=\mu^2, \omega \cd q=0)$ are zero. \item The values of
all bare parameters and counterterms for $l\leq N$ are determined
from successive calculations within the $1,2,\dots N$ Fock space
truncations.
\end{itemize}
}

\section{Applications to the N=2 Fock space truncation} \label{QED}
In order to show simple but nevertheless meaningful applications
of the general renormalization scheme developed above, we
consider, in a first step, the Yukawa model. The LF Hamiltonian
including PV fields  has been derived in Sec.~\ref{clfd}. We
shall then extend our results to QED which is very similar to the
Yukawa model for the $N=2$ Fock space truncation. To simplify
notations, we will omit in the following  the superscript $(N)$ at
each vertex function.

\subsection{Yukawa model}
\label{Yukmodel}
\subsubsection{Solution of the eigenstate equation} \label{eigenst}
The solution of the eigenstate equation~(\ref{eq3})
within the $N=2$  Fock space  truncation has already been found
in Ref.~\cite{kms_04}. Sharp cut-offs
imposed on the longitudinal and transverse components of particle
momenta were used to regularize the amplitudes. We revisit here
the same problem, but apply the
PV regularization method (with one PV fermion and one PV boson),
as advocated in Sec.~\ref{clfd}. Besides that, we shall apply and
test the new renormalization scheme proposed in Sec.~\ref{renor}.

The use of the PV regularization extends Fock space: instead of
one Fock component for the one-body sector we have now two
components corresponding to the physical and PV fermions, while
the two-body sector is described by four components related to the
states either with both physical particles, or with the physical
fermion plus the PV boson, or with the PV fermion plus the
physical boson, or, finally, with both PV particles. The extension
of Fock space makes the computational analysis more cumbersome,
but this is compensated by the simplification of the equations for
the Fock components due to the disappearance of the contact terms,
as well as the absence of spurious $\omega$-dependent
contributions to the fermion self-energy~\cite{kms_07}.

 To incorporate the PV particles into the state vector, we will   supply the
vertex functions, as well as the particle momenta and masses, with
the indices $i$ and $j$ which relate, respectively, to fermions
and bosons. Each index is either $0$ for a physical particle or
$1$ for a PV one. All particle momenta are on their mass shells:
$$
k_{i}^2=m_{i}^2,\quad k_{j}^2=\mu_{j}^2,
$$
with $m_{0}=m$ and $\mu_{0}=\mu$ being the physical particle
masses.

Following Eq.~(\ref{onetwo}), we decompose the vertex functions in
invariant amplitudes:
\begin{subequations} \label{gm12t}
\begin{equation}
\label{gm1inv}
\bar{u}(p_{1i})\Gamma_{1}^{i}u(p)  =
(m_{i}^2-M^2)\psi_{1}^{i}\bar{u}(p_{1i})u(p) ,
\end{equation}
\begin{equation}
\label{gm2inv}\bar{u}(k_{1i})\Gamma_{2}^{ij}u(p)  =
\bar{u}(k_{1i})\left[b_{1}^{ij}+b_{2}^{ij} {\displaystyle
\frac{m{ \sla \omega}}{\omega\cd p}}\right]u(p).
\end{equation}
\end{subequations}
Since the fermion momenta for the one- and two-body vertices are
different, we denote them by different letters, namely, by
$p_{1i}$ for the one-body vertex and by $k_{1i}$ for the two-body
one. The boson four-momentum is $k_{2j}$.  Here $M$ is a temporary
notation for the physical fermion mass (in the end of the
calculation we will take the limit $M\to m$), $k_{1i}^2 =
p_{1i}^2=m_{i}^2$, $\psi_{1}^{i}$ are constants (i.e. they do not
depend on particle momenta), while the invariant functions
$b_{1}^{ij}$ and $b_{2}^{ij}$ may have momentum
dependence.\footnote{In the two-body approximation, as we will
see, they reduce to constants, but this is no more the case in
higher order approximations.} We introduce the standard LFD
variables
\begin{equation}
\label{lf} x=\frac{\omega\cd k_{2j}}{\omega\cd p},\quad { \bf
R}_{\perp}={\bf k}_{2j\perp}-x{\bf p}_{\perp}
\end{equation}
which are, respectively, the longitudinal and transverse (with respect to the
LF plane orientation)
components of the bosonic momentum. Note that the square of the two-dimensional vector
${\bf R}_{\perp}$ is an invariant: $R_{\perp}^2=-(k_{2j}-xp)^2$.

\begin{figure}[ht!]
\begin{center}
\includegraphics[width=20pc]{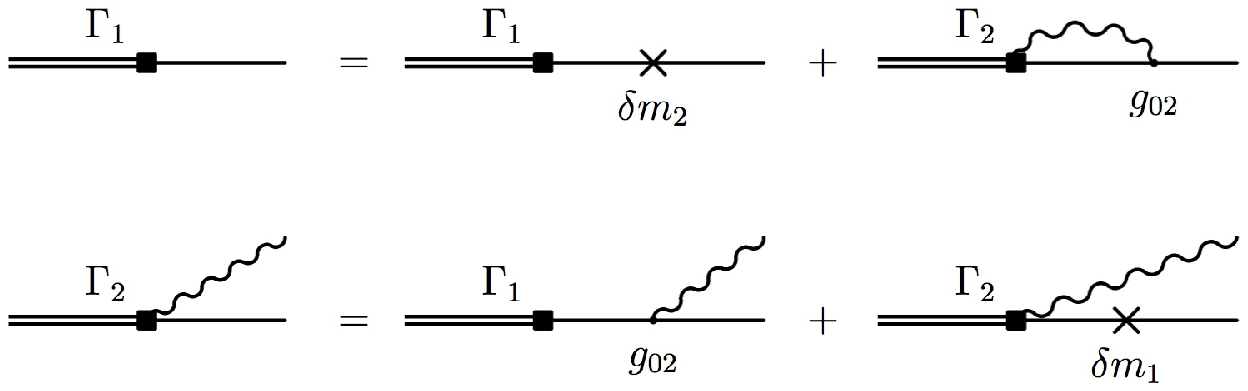}
\caption{System of equations for the vertex functions in the
two-body approximation.}\label{fig1}
\end{center}
\end{figure}

The system of coupled equations for the vertex functions in the
two-body truncated Fock space is shown graphically in
Fig.~\ref{fig1}.
Since the Hamiltonian~(\ref{hamPV}) does not include contact
terms, this system is much simpler than its analog considered in
Ref.~\cite{kms_04}. For simplicity, we have not drawn the lines associated with the spurion (see discussion in Appendix \ref{appB}). Besides that, we do not need to introduce any
specific counterterms which explicitly depend on the LF plane
orientation. These are evident merits of the PV regularization.
The amplitudes of the LF diagrams are calculated according to the
graph technique rules (see Ref.~\cite{cdkm} and Appendix
\ref{appB}). We denote  the intermediate fermion four-momenta in
the one- and two-body states as $p'_{1i'}$ and $k'_{1i'}$,
respectively. The intermediate boson four-momentum is
$k'_{2j'}$. The values of the intermediate particle momenta are
defined from the conservation laws~(\ref{k1n}) in the vertices,
taken for $n=1$ and $n=2$.

The system of equations reads
\begin{subequations} \label{eq2t}
\begin{eqnarray}
 \bar{u}(p_{1i})\Gamma_{1}^{i}u(p) &=&
\bar{u}(p_{1i})\left(V_{1}+V_{2}\right)u(p) , \label{eq2b1}\\
\bar{u}(k_{1i})\Gamma_{2}^{ij}u(p) &= &\bar{u}(k_{1i})V_{3}u(p) ,\label{eq2b2}
\end{eqnarray}
\end{subequations}
where
\begin{subequations} \label{V22t}
\begin{eqnarray}
\label{V12b} V_{1} & = & \delta m_{2}
\sum_{i'}(-1)^{i'}\frac{{\sla p_{1i'}'}+m_{i'}}{m^2_{i'}-M^2}
\Gamma_{1}^{i'},  \\
 V_{2}& = & g_{02}\int \frac{d^2R'_{\perp}}{(2\pi)^3}\int_{0}^{1}
 \frac{dx'}{2x'(1-x')} \label{V22b}\\
&&\times\left[\sum_{i',j'}(-1)^{i'+j'} \frac{{\sla k}'_{1i'}+m_{i'}} {s_{12}^{i'j'}-M^2}
\Gamma_{2}^{i'j'}\right] ,\nonumber\\
\label{V32b} V_{3} & = & g_{02}\sum_{i'}(-1)^{i'} \frac{{\sla
p'_{1i'}}+m_{i'}}{m^2_{i'}-M^2} \Gamma_{1}^{i'},
\end{eqnarray}
\end{subequations}
and $s_{12}^{i'j'}=(k'_{1i'}+k'_{2j'})^2$. We took into account
that $\delta m_1=0$. Necessary kinematical relations used for
deriving Eqs.~(\ref{V22t}) are given in Appendix \ref{kin1}. We
introduced the  integration variables ${\bf R}'_{\perp}$ and $x'$
related to the intermediate boson momentum $k'_{2j'}$ in full
analogy with Eqs.~(\ref{lf}).
 The summations in Eqs.~(\ref{V22t}) run through 0 to 1 in each index.
 The system of equations~(\ref{eq2t}) must be solved in the limit $M\to m$.

We have four unknown  functions $b_{1}^{ij}$, four unknown
functions $b_{2}^{ij}$ and two unknown constants $\psi_{1}^{i}$.
So, we have to deal with a system of $10\times 10$ linear integral
equations. However, Eqs.~(\ref{eq2b2}) and~(\ref{V32b}) allow to
express easily $\Gamma_{2}^{ij}$ through $\Gamma_{1}^{i}$. Since
$\Gamma_{1}^{i}$ are constants, it follows from these equations
that $b_{1}^{ij}$ and $b_{2}^{ij}$ are constants too and they
depend neither on $i$, nor on $j$. Then, due to the fact that the
vertex functions come into all equations, being sandwiched with
on-mass-shell bispinors, we may simply substitute
$\Gamma_{2}^{ij}$ everywhere by the quantity $V_{3}$,
Eq.~(\ref{V32b}).  It thus follows that we can rewrite
Eq.~(\ref{V22b}) in the form, in the limit $M \to m$
\begin{equation}
\label{V22bs} V_{2} = -g_{02}\left.\bar{\Sigma}(p)\right
\vert_{p^2=M^2} V_{3},
\end{equation}
where
\begin{eqnarray}
\label{sigma} \bar{\Sigma}(p)&=&-\int
\frac{d^2R'_{\perp}}{(2\pi)^3}\int_{0}^{1} \frac{dx'}{2x'(1-x')}\nonumber \\
&&\times\left[\sum_{i',j'}(-1)^{i'+j'} \frac{{\sla k_{1i'}'}+m_{i'}}
{s_{12}^{i'j'}-p^2} \right]
\end{eqnarray}
which is nothing else than the PV-regularized fermion two-body
self-energy (apart from the coupling constant). After
integration, it depends on the four-momentum $p$ only. We will use
the following decomposition:
\begin{equation}
\label{sigmainv} \bar{\Sigma}(p)\equiv {\cal A}(p^2)+{\cal
B}(p^2)\frac{\sla p}{m} .
\end{equation}
The explicit form of the functions ${\cal A}$ and ${\cal B}$ for
arbitrary values of $p^2$ can be found in Ref.~\cite{kms_07}.

Now, substituting Eq.~(\ref{V32b}) into Eq.~(\ref{V22bs}) and then
into Eq.~(\ref{eq2b1}), and using that $\Gamma_1^{i}=(m^2_{i}-M^2)\psi_1^i$  is
proportional to the unity matrix, we turn the latter equation into a system of two
linear matrix equations for $\psi_{1}^{0}$ and $\psi_{1}^{1}$.
Multiplying these equations by $\bar{u}(p)$ to the right and by
${u}(p_{1i})$ to the left, summarizing over the fermion polarizations, and
taking the trace, we arrive at a system of two linear
homogeneous equations for the one-body vertices. This system
reads
\begin{equation}
\label{ur2b} \sum_{i'}c^{ii'}\psi_{1}^{i'}=0
\end{equation}
with
\begin{eqnarray*}
c^{00} & = & 8m^2\left\{\delta m_{2}-g_{02}^2(A+B)\right\},\\
c^{01} & = & -2(m+m_1)^2\left\{\delta m_{2}-g_{02}^2(A+B)\right\},\\
c^{10} & = & 2(m+m_1)^2\left\{\delta m_{2}-g_{02}^2(A+B)\right\},\\
c^{11} & = & \frac{(m+m_1)^2}{m^2}\left\{m^3-mm_1\left(m_1+2\delta
m_{2}\right) \right. \\
&&+\left.g_{02}^2\left[2Amm_1+B(m^2+m_1^2)\right]\right\} ,
\end{eqnarray*}
where $A={\cal A}(m^2)$, $B={\cal B}(m^2)$. Equating the
determinant of the system to zero, we obtain a quadratic equation
for $\delta m_{2}$ with the solution
\begin{equation}
\label{dm(2)} \delta m_{2}=g_{02}^2(A+B).
\end{equation}
The second root is rejected because it does not disappear at
$g_{02}=0$. Substituting Eq.~(\ref{dm(2)}) into any of the two
equations~(\ref{ur2b}) we find
\begin{equation}
\label{psi01} \psi_{1}^{0}=a_{1},\quad \psi_{1}^{1}=0,
\end{equation}
where $a_{1}$ is an arbitrary constant. Substituting this solution
into Eq.~(\ref{V32b}), then into Eq.~(\ref{eq2b2}), and comparing
the result with the r.-h.s. of Eq.~(\ref{gm2inv}), we easily
obtain
\begin{equation}
\label{b12} b_{1}^{ij}=2mg_{02}a_{1},\quad b_{2}^{ij}=0.
\end{equation}

It is interesting to compare the solution~(\ref{psi01})
and~(\ref{b12}) for the one-  and two-body vertex functions with
that found in Ref.~\cite{kms_04}. Due to the extension of the Fock
space basis, the vertex functions considered here have more
components, but both solutions possess nevertheless the same main
features. Firstly, they are constants, i.e. momentum independent.
Secondly, the one-body vertex has only one component:  the
one-body wave function of the PV fermion, $\psi_{1}^1$, vanishes
identically, while the physical component $\psi_{1}^0$ is not
fixed and must be computed from the normalization condition for
the state vector. Thirdly, the two-body vertex has no
$\omega$-dependent part, since the components $b_{2}^{ij}$ are
zero. Finally, the form of the solution~(\ref{b12}) is the same as
that from Ref.~\cite{kms_04}, apart from the coupling constant
which is now $g_{02}$ instead of $g$. The same is true for the
MC~(\ref{dm(2)}).

\subsubsection{Normalization of the state vector}
\label{svn}
The Fock components found above must be properly normalized.
 As we have already explained, the formulas~(\ref{normwf2}) and~(\ref{intGamma})
 for the "partial" normalization integrals must be modified, since we have to take
into account the sectors which contain PV particles. This is done
very easily. One should simply sum over all possible two-body
states, keeping in mind that each PV particle brings the factor
$(-1)$ to the norm. The contribution of the one-body state is thus
\begin{eqnarray}
I_1 & = & \sum_{i}(-1)^{i}\frac{\mbox{Tr}\{({\sla
p}+m)({\sla
p}_{1i}+m_i)\}}{2}\left(\psi_1^i\right)^2\nonumber \\
&=&4m^2\left(\psi_1^0\right)^2=4m^2a_1^2\label{norm2b1}.
\end{eqnarray}
We used here Eqs.~(\ref{Ap1a}) and~(\ref{Ap2}) at $M=m$ from
Appendix \ref{kin1}. The norm of the two-body state reads
\begin{eqnarray}
I_2&=&\frac{1}{2(2\pi)^3}\int d^2R_{\perp} \int_0^1
\frac{dx}{2x(1-x)}\sum_{ij}(-1)^{i+j}\nonumber \\
\label{norm2b2}
&&\times\frac{\left(b_1^{ij}\right)^2\mbox{Tr}\{({\sla p}+m)({\sla
k}_{1i}+m_i)\}}{(s_{12}^{ij}-m^2)^2},
\end{eqnarray}
where
$$
s_{12}^{ij}=\frac{R_{\perp}^2+\mu_j^2}{x}+\frac{R_{\perp}^2+m_i^2}{1-x}.
$$
Substituting $b_1^{ij}=2m{g}_{02}a_1$ into Eq.~(\ref{norm2b2}),
calculating the trace and using Eqs.~(\ref{Ap1c}), (\ref{Ap4}),
and~(\ref{Ap5}), we obtain
$$
I_2=\frac{m^2a_1^2{g}_{02}^2}{2\pi^2} \int_0^{\infty}
dR_{\perp}\,R_{\perp}\int_0^1 dx\,x\sum_{i,j}(-1)^{i+j}
$$
\begin{equation}
\label{norm2b3}
\times
\frac{R_{\perp}^2+[(1-x)m+m_i]^2}{[R_{\perp}^2+(1-x)\mu_j^2+xm_i^2-x(1-x)m^2]^2}.
\end{equation}
Note that in order to get a finite result for $I_2$, the bosonic PV regularization is
enough. For this reason, one can retain in the sum over $i$
the term with $i=0$ only\footnote{Neglecting
the term with $i=1$ brings corrections of relative order $(m/m_1)^2\log(m_1/m)$,
which tend to zero as the PV fermion mass $m_1$ increases to infinity.} :
\begin{equation}
\label{norm2b4} I_2=4m^2a_1^2{g}_{02}^2\,J_2,
\end{equation}
where
\begin{eqnarray}
\label{norm2b4a}
J_2 & = & \frac{1}{8\pi^2}\int_0^{\infty}
dR_{\perp}\,R_{\perp}\int_0^1 dx\,x\nonumber \\
& &\times \sum_{j}
\frac{(-1)^{j}[R_{\perp}^2+(2-x)^2m^2]}{[R_{\perp}^2+(1-x)\mu_j^2+x^2m^2]^2}.
\end{eqnarray}
From the normalization condition~(\ref{normwf1}) in the two-body
approximation one gets $I_1+I_2=1$ and, hence,
\begin{equation} \label{Z2g0}
a_1^2=\frac{1}{4m^2(1+g_{02}^2 J_2)}.
\end{equation}
The integral $J_2$ diverges logarithmically when the PV boson mass $\mu_1$ tends to infinity:
\begin{equation}
\label{J2div}
J_2=\frac{1}{16\pi^2}\log\frac{\mu_1}{m}+f\left(\frac{\mu}{m}\right),
\end{equation}
where $f$ is a finite function independent of $\mu_1$. Terms vanishing
in the limit $\mu_1\to\infty$ are neglected. Note that if $\mu_1$ is large enough,
$J_2$ is positive.

The normalized one- and two-body components of the vertex functions are thus
\begin{subequations} \label{onetwonorm}
\begin{eqnarray}
\label{onebnorm} \psi_1^0=\frac{1}{2m\sqrt{1+g_{02}^2J_2}},\quad
\psi_1^1=0,\\
\label{twobnorm}
b_1^{ij}=\frac{g_{02}}{\sqrt{1+g_{02}^2J_2}},\quad b_2^{ij}=0.
\end{eqnarray}
\end{subequations}
%
\subsubsection{Determination of the internal  bare coupling constant} \label{ABCC}
According to our renormalization conditions detailed in
Sec.~\ref{renorcond}, we calculate the internal BCC $g_{02}$ from
the requirement that the $\omega$-independent part of the two-body
vertex function for physical particles (i.e. the component
$b_1^{00}$), taken  at  $s=m^2$, is identified with the physical
coupling constant $g$. Since in the two-body Fock space truncation
$b_1^{ij}$ are constants, we immediately get
\begin{equation} \label{a2}
b_1^{00}(s=m^2)\equiv b_1^{00} =
\frac{g_{02}}{\sqrt{1+g_{02}^2J_2}} = g.
\end{equation}
It is a well defined, non-perturbative, condition which is very convenient to
impose in any numerical calculation.

From Eq.~(\ref{a2}) it follows
\begin{equation} \label{g0b22}
g_{02}^2=\frac{g^2}{1-g^2J_2}.
\end{equation}
The final form of the normalized (and renormalized) solution for
the vertex function components becomes
\begin{subequations} \label{onetworenor}
\begin{eqnarray}
\label{onebrenorm}
\psi_1^0=\frac{\sqrt{1-g^2J_2}}{2m},\quad \psi_1^1=0,\\
\label{twobrenorm}
b_1^{ij}=g,\quad b_2^{ij}=0.
\end{eqnarray}
\end{subequations}
The one- and two-body contributions to the norm of the state vector are
\begin{equation}
\label{I1I2ren}
I_1=1-g^2J_2,\quad I_2=g^2J_2.
\end{equation}

As we said above, if $\mu_1$ increases, the quantity $J_2$
increases too. At some value of $\mu_1$ we inevitably meet the
condition $g^2J_2=1$, leading to a pole on the r.-h.s. of
Eq.~(\ref{g0b22}), analogous to the well-known Landau pole in QED.
Further increase of $\mu_1$ makes $g_{02}^2$ negative and $g_{02}$
purely imaginary [as well as $\psi_1^0$, Eq.~(\ref{onebnorm})]. At
the same time, the one- and two-body norms $I_1$ and $I_2$ become
infinitely large,  but have opposite signs: $I_1$ is negative,
while $I_2$ is positive. So one can not get rid  at this level of
the regularization parameters by taking the limit $\mu_1\to
\infty$ without formal contradiction. A possible way out is as
follows. The PV masses play an auxiliary role and appear in
intermediate calculations only, while physical observables must be
independent of them (the BCC $g_{02}$ is not an observable!). We
do not give any physical interpretation to intermediate results
found with finite PV masses. Below we will treat $J_2$ as being a
{\em finite} quantity satisfying the inequality $1-g^2J_2>0$
(formally, this can be done by proper adjustment of the value of
$\mu_1$). Then, we express the calculated Fock components through
the physical coupling constant, normalize the state vector,
calculate the observables, and after that go over to the limit of
infinite PV masses in the {\em final result}. We will demonstrate
below  how this scheme works in practice.

\subsubsection{Determination of the external  bare coupling constant}
\label{ffsYuk}
To determine the external BCC one has to consider the 3PGF, denoted by  $G$,
which is represented, in the two-body approximation, by a sum of
the two contributions shown in Fig.~\ref{ff2}. Similarly to the
two-body vertex function~(\ref{gm2inv}), the 3PGF can be
decomposed in invariant amplitudes:
\begin{figure}[btph]
\begin{center}
\includegraphics[width=20pc]{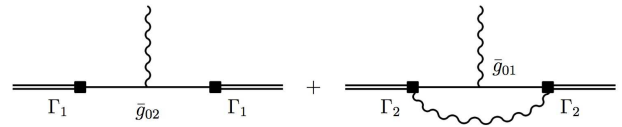}
\caption{Fermion-boson 3PGF in the two-body approximation.
\label{ff2}}
\end{center}
\end{figure}
\begin{equation}
\label{3PGFY}
\bar{u}(p')Gu(p)=g\bar{u}(p')\left(F+B_{\omega}\frac{m\sla
\omega}{\omega\cd p}\right)u(p).
\end{equation}
The invariant functions $F$ and $B_{\omega}$ (scalar form
factors)  depend, under the condition $\omega\cd q=0$, on
$q^2\equiv (p'-p)^2$. Note that the general decomposition
(\ref{3PGFY}) in the Yukawa model
 is valid in any approximation, since the number of independent invariant amplitudes
is completely determined by particle spins and by symmetries of
the interaction. The external BCC is found from the requirement
\begin{equation}
\label{ren3PGFY}
\bar{u}(p')Gu(p)=g\bar{u}(p')u(p)\quad \mbox{at}\quad q^2=\mu^2.
\end{equation}
Hence, we must have $F(q^2=\mu^2)=1$ and
$B_{\omega}(q^2=\mu^2)=0$. In exact calculations, as well as in a
given order of perturbation theory, we would indeed get
$B_{\omega}\equiv 0$, because physical amplitudes can not depend
on the LF plane orientation. In approximate non-perturbative
calculations however it may turn out that $B_{\omega}$ does not
disappear, even at $q^2=\mu^2$. Then, according to our
renormalization prescriptions, one should add to the LF
Hamiltonian the counterterm
$$
Z_3\bar{\psi}\frac{m\sla \omega}{i\omega\cd
\partial}\psi\varphi \ .
$$
The structure of the
counterterm is the same in all approximations, while the constant
$Z_3$ must depend on the Fock sector, in complete analogy with the BCC's. We therefore have to calculate both $Z_{3 l}$ (for internal coupling) and $\bar Z_{3l}$ (for external coupling). In the particular case discussed here, i.e for the two-body
Fock space truncation with the use of the PV regularization (with
one PV boson and one PV fermion), we find $B_{\omega} = 0$. We therefore get  $\bar Z_{31}=\bar Z_{32}=Z_{31}=Z_{32}=0$.

We will now concentrate
on the calculation of the physical scalar form factor $F$. It
can be extracted from the 3PGF by the relation
\begin{equation}
\label{3PGFYF1}
gF=\frac{1}{2q^2}\,\mbox{Tr}\left[({\sla p}'+m)G({\sla
p}+m)\left(\frac{m{\sla \omega}}{\omega\cd p}-1\right)\right].
\end{equation}
First, we calculate $F$, as a function of $q^2$, for physical
values $q^2\leq 0$ and then find its analytical continuation to
the non-physical point $q^2=\mu^2$.

In the two-body approximation, the 3PGF can be represented as
\begin{equation}
\label{vert3PGF} G\equiv \bar{g}_{02}\bar G_1+\bar{g}_{01}\bar
G_2,
\end{equation}
where $\bar G_{1,2}$ denote the contributions (amputated from the
external BCC's) to the full 3PGF from the one- and two-body
sectors, respectively. Applying the CLFD graph technique rules to
the diagrams shown in Fig.~\ref{ff2}, we have
\begin{subequations}
\begin{eqnarray}
\bar G_1&=&\sum_{i,i'}(-1)^{i+i'}
\psi_1^i\psi_1^{i'}\left[({\sla p}'+{\sla \omega}\tau_1'+m_{i'})\right.
\nonumber \\
\label{oneb3PGF} &&\times\left.({\sla
p}+{\sla \omega}\tau_1+m_i)\right], \\
\bar G_2 & = &\frac{1}{(2\pi)^3}\int
d^2R_{\perp}\int_0^1\frac{dx}{2x}\sum_{i,i',j}(-1)^{i+i'+j}
\nonumber \\
&&\times \frac{\bar{\Gamma}_2^{i'j}({\sla p}'-{\sla k}_{2j}+{\sla
\omega}\tau_2'+m_{i'})}
{(p'-k_{2j})^2-m_{i'}^2}\nonumber \\
\label{twob3PGF} &&\times \frac{({\sla p}-{\sla k}_{2j}+{\sla
\omega}\tau_2+m_{i})\Gamma_2^{ij}} {(p-k_{2j})^2-m_{i}^2},
\end{eqnarray}
\end{subequations}
where the variables $x$ and ${\bf R}_{\perp}$ are defined by
Eqs.~(\ref{lf}), and $\tau$'s are given by Eqs.~(\ref{Ap2}),
(\ref{Ap5}), and~(\ref{Ap5a}) taken for $M=m$. Kinematical
relations needed to express the scalar products of the
four-momenta through $x$ and ${\bf R}_{\perp}$ can be found in
Appendix~\ref{kin2}. To make the integral in Eq.~(\ref{twob3PGF})
convergent, it is enough to regularize it by the bosonic PV
subtraction only, as we did for the calculation of $J_2$ in
Eq.~(\ref{norm2b4a}). We can thus neglect in the sum the terms
with either $i$ or $i'$ being 1. According to the
solution~(\ref{twobrenorm}) for the two-body component of the
state vector, $\Gamma_2^{ij}=b_1^{ij}=g$.  Using the
solution~(\ref{onebrenorm}) for the one-body component,
substituting Eqs.~(\ref{oneb3PGF}) and~(\ref{twob3PGF}) into
Eq.~(\ref{vert3PGF}) and then into Eq.~(\ref{3PGFYF1}), we find
\begin{equation}
\label{F1q3PGF}g F=\bar{g}_{02}{\cal F}^{(1)}+\bar{g}_{01}{\cal
F}^{(2)}(q^2),
\end{equation}
where
\begin{subequations}
\begin{widetext}
\begin{eqnarray}
\label{F113PGF} {\cal F}^{(1)}&=&1-g^2J_2,\\
\label{F12q3PGF} {\cal F}^{(2)}(q^2)&=&\frac{g^2}{16\pi^3}\int
d^2R_{\perp}\int_0^1 dx\sum_{j} \frac{(-1)^jx[2({\bf
R}_{\perp}{\bg \Delta})^2/\Delta^2-R_{\perp}^2-x{\bf
R}_{\perp}{\bg
\Delta}+(3x^2-8x+4)m^2]}{(1-x)[R_{\perp}^2+(1-x)\mu_j^2+x^2m^2]
[({\bf
R}_{\perp}-x{\bg \Delta})^2+(1-x)\mu_j^2+x^2m^2]}\nonumber \\
\end{eqnarray}
\end{widetext}
\end{subequations}
with ${\bg \Delta}={\bf q}_{\perp}$ and $\Delta^2=-q^2$. The
functions ${\cal F}^{(1)},{\cal F}^{(2)}$ determine the
contributions to the scalar form factor $F$ from the one- and
two-body Fock sectors. The integration over $d^2R_{\perp}$ in
Eq.~(\ref{F12q3PGF}) can be performed analytically and leads to
the result
$$
{\cal F}^{(2)}(q^2)=\frac{g^2}{4\pi^2\sqrt{-q^2}}\int_0^1dx\sum_j(-1)^j
$$
\begin{equation}
\label{F12q3PGFx}\times \frac{[4(1-x)m^2+\mu_j^2]}{r_j(q^2)}\,
\log\left[\frac{x\sqrt{-q^2}+r_j(q^2)}{2\sqrt{x^2m^2+(1-x)\mu_j^2}}\right],
\end{equation}
where
$$
r_j(q^2)=\sqrt{x^2(4m^2-q^2)+4(1-x)\mu_j^2}.
$$

\begin{figure}[btph]
\begin{center}
\includegraphics[width=16pc]{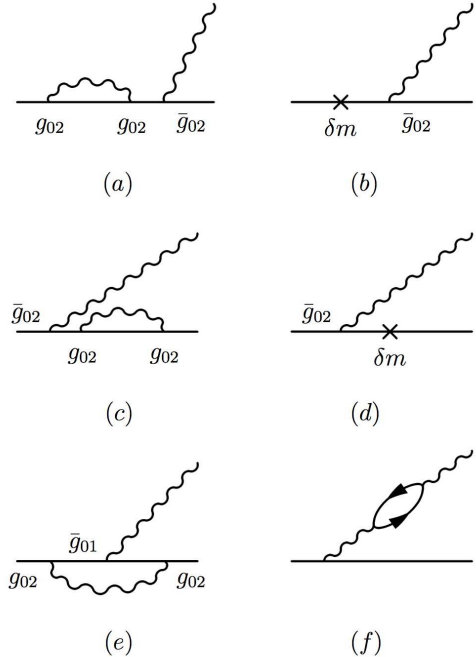}
\caption{Second order perturbative corrections to the
internal and external BCC's. \label{RC2}}
\end{center}
\end{figure}

According to Eq.~(\ref{g01}), $\bar{g}_{01}=g$, and
Eq.~(\ref{F1q3PGF}) contains one unknown parameter, namely, the
external BCC $\bar{g}_{02}$
which is found from the renormalization condition
$F(q^2=\mu^2)=1$. We thus get
\begin{equation}
\label{g021} g=\bar{g}_{02}(1-g^2J_2)+g{\cal F}^{(2)}(\mu^2),
\end{equation}
where $J_2$ is defined by Eq.~(\ref{J2div}) and
$$
{\cal F}^{(2)}(\mu^2)=\frac{g^2}{4\pi^2\mu}\int_0^1dx\sum_j(-1)^j
$$
\begin{equation}
\label{F2mu}\times
\frac{[4(1-x)m^2+\mu_j^2]}{r_j(\mu^2)}\arctan\left[\frac{\mu
x}{r_j(\mu^2)}\right].
\end{equation}
The asymptotic value of ${\cal F}^{(2)}(\mu^2)$ at
$\mu_1\to\infty$ is
\begin{equation}
\label{F2mulim} {\cal
F}^{(2)}(\mu^2)=-\frac{g^2}{8\pi^2}\log\frac{\mu_1}{m}+g^2f_1\left(\frac{\mu}{m}\right),
\end{equation}
where $f_1$ is a function of the ratio $\mu/m$, finite at
$\mu_1\to\infty$. Substituting Eqs.~(\ref{F2mulim})
and~(\ref{J2div}) into Eq.~(\ref{g021}), we find for the external
BCC:
\begin{equation}
\label{g022} \bar{g}_{02}=g\,\frac{{\displaystyle
1+\frac{g^2}{8\pi^2}\log\frac{\mu_1}{m}-g^2f_1\left(\frac{\mu}{m}\right)}}
{{\displaystyle
1-\frac{g^2}{16\pi^2}\log\frac{\mu_1}{m}-g^2f\left(\frac{\mu}{m}\right)}}.
\end{equation}
The internal and external BCC's given by Eqs.~(\ref{g0b22})
and~(\ref{g022}), respectively, differ from each other. We have
already mentioned in Sec.~\ref{BCCg} that they indeed should not
coincide. We can illustrate the origin of this difference in a
very clear form, by analyzing contributions to the internal and
external BCC's from various LF diagrams taken into account in our
calculations. Since $g_{02}$ and $\bar{g}_{02}$ differ already at
order $g^2$ of their perturbative expansions, it is enough, 
to clarify the situation, to consider the lowest order perturbative
corrections to BCC's. They are indicated in Fig.~\ref{RC2} where the outgoing wavy line corresponds to an external boson. As it
has been pointed out above, only the diagrams (a) and (b) are
incorporated when calculating the internal BCC $g_{02}$ in the
two-body approximation. Concerning the external BCC
$\bar{g}_{02}$, the diagrams (c), (d), and (e) are included in
addition, though they formally correspond to a three-body state
admixture,  with the external field. In other words, our
calculation of $\bar{g}_{02}$ is effectively performed to higher
order approximation, than that of $g_{02}$. One may expect that
increasing the number of Fock components involved into
calculations will reduce the difference between the internal and
external BCC's.

The diagram (f) in Fig.~\ref{RC2}, containing the
fermion-antifermion pair intermediate state, does not contribute,
in the two-body approximation, neither to $g_{02}$ nor to
$\bar{g}_{02}$. For this reason, both BCC's, even being
expanded in powers of $g$ up to terms of order $g^2$, differ from the
BCC found in the second order of perturbation theory.

As follows from Eqs.~(\ref{F1q3PGF}) and~(\ref{g021}), the scalar
form factor $F$ is given by
\begin{equation}
\label{FYUK} F=1+{\cal F}^{(2)}(q^2)-{\cal F}^{(2)}(\mu^2).
\end{equation}
It is easy to see that the expression on the r.-h.s. is finite in
the limit $\mu_1\to\infty$, though each of the functions ${\cal
F}^{(2)}$ diverges logarithmically. Eq.~(\ref{FYUK}) coincides
with the result of the 3PGF renormalization in the second order
of perturbation theory, in spite of the fact that we did not make
any expansion in powers of the coupling constant.

\subsection{Application to QED}
\subsubsection{Determination of the state vector}
The Yukawa model considered in the previous section is a good
example of how the  general renormalization scheme developed in
this article should be understood. We will now apply the method to
QED in order to address the case of a realistic physical theory.
The perturbative approach to QED was able to reproduce multiple
experimental data with excellent precision. Applying the
non-perturbative scheme to the same object gives us a possibility
to test our results and to reveal distinctly its main differences,
as compared to perturbation theory.

From a purely technical point of view, QED is more complicated
than the Yukawa model, at least, in the following three aspects.
Firstly, the structure of the interaction Hamiltonian in QED is
more involved than that in the case of scalar bosons. The
consideration of QED within CLFD, performed in Ref.~\cite{kms_04},
showed that the corresponding Hamiltonian (obtained without
introducing PV fields) contained specific contact terms, different
from those which appear in the Yukawa model. It is not yet clear
whether PV fields can kill the contact terms in this case.
 Secondly, many-body wave functions in QED have more spin components
than their counterparts in the Yukawa model. Thirdly, intermediate
calculations in QED essentially depend on the gauge condition to
constrain the electromagnetic field, while the question how the
Fock space truncation affects gauge invariance is still
opened.

The difficulties of treating QED in the framework of CLFD,
itemized above, are however absent for the two-body Fock space
truncation. In this approximation, QED and the Yukawa model are
very close to each other. We repeat below the procedure of finding
the state vector and its renormalization, detailed in
Sec.~\ref{Yukmodel}, for QED in the two-body truncated Fock space.
Because of similarities between QED and the Yukawa model, we will
not expose all the steps, but concentrate on pointing out the
differences and demonstrating the main results of our analysis.

The general structure of the two-body electron-photon vertex
function in QED has been extensively studied in
Ref.~\cite{kms_04}. For the case of the simple two-body Fock space
truncation, the interaction Hamiltonian in the Feynman gauge is
very similar to that in the Yukawa model, with the scalar vertex
being replaced by the electromagnetic one. Provided the PV
regularization is used, with one PV boson and one PV fermion, the
fermion self-energy does not depend on the  LF plane orientation.

According to Ref.~\cite{kms_04}, the state vector has the
following structure:
\begin{subequations}
\begin{equation}
\bar{u}(p_{1i})\Gamma_{1}^{i}u(p)  =  (m_{i}^2-M^2)\psi_{1}^{i}
\bar{u}(p_{1i})u(p) ,\label{QEDinv1}
\end{equation}
\begin{multline}
 \bar{u}(k_{1i})\left[\Gamma_{2}^{ij,\nu}e_\nu^\lambda (k_{2j})\right]
 u(p) \\
\label{QEDinv2}
 = \bar{u}(k_{1i})\left[b_{1}^{ij} \gamma^\nu+b_{2}^{ij} {\displaystyle
\frac{m{ \omega^\nu}}{\omega\cd p}}\right]e_\nu^\lambda (k_{2j}) u(p) ,
\end{multline}
\end{subequations}
where $e_\nu^\lambda (k_{2j}) $ is the photon polarization vector.
The scalar functions $\psi_{1}^{i}$, $b_{1}^{ij}$, and
$b_{2}^{ij}$ differ, generally speaking,  from those entering
Eqs.~(\ref{gm12t}). The decomposition~(\ref{QEDinv2}) with the two
matrix components is valid only within the two-body Fock space
truncation. The number of independent components of the two-body
vertex function in QED depends on the gauge. If the Feynman gauge
is chosen, there are eight independent components in the general case.
Under the two-body approximation, six of them identically turn
into zero.

The system of equations for the QED vertex functions is very close
to that for the Yukawa model, shown in Fig.~\ref{fig1} and given
analytically by Eqs.~(\ref{eq2t}) and~(\ref{V22t}).  Small changes
are required, caused by the vector character of the photon.
Namely, one has to substitute $g_{02}\to e_{02}\gamma^{\rho}$,
$\Gamma_2^{ij}\to \Gamma_2^{ij,\nu}$, and $\bar{\Sigma}(p)\to
\bar{\Sigma}_{\rho\nu}(p)$, where, in the Feynman gauge,
\begin{equation}
\label{Srhonu}
\bar{\Sigma}_{\rho\nu}(p)=-g_{\rho\nu}\left[{\cal A}(p^2)+
{\cal B}(p^2)\frac{\sla p}{m}\right].
\end{equation}
The scalar functions ${\cal A}$ and ${\cal B}$ here differ from
those in Eq.~(\ref{sigmainv}). They were calculated (for the
longitudinal and transversal LFD cutoffs) in Ref.~\cite{kms_04}.
We do not need to know their explicit form in the following.

The technical procedure to solve the system of equations for the
vertex functions is exactly the same as in Sec.~\ref{eigenst}. The
solution looks as follows:
\begin{equation}
\label{dm(2)v} \delta m_{2}=2e_{02}^2(2A-B),
\end{equation}
with $A={\cal A}(m^2)$, $B={\cal B}(m^2)$, and
\begin{subequations}
\begin{eqnarray}
\label{psi01v} \psi_{1}^{0}=a_{1},\quad \psi_{1}^{1}=0,\\
\label{b12v} b_{1}^{ij}=2me_{02}a_{1},\quad b_{2}^{ij}=0.
\end{eqnarray}
\end{subequations}
Eqs.~(\ref{psi01v}) and~(\ref{b12v}) coincide in form with
Eqs.~(\ref{psi01}) and~(\ref{b12}), respectively. The constant
$a_1$ is found from the normalization condition.

\subsubsection{Normalization of the state vector}
\label{normQED}
The calculation of the one- and two-body normalization integrals
is analogous to that for the Yukawa model. The one-body integral
is exactly the same as in Eq.~(\ref{norm2b1}), since it is not
sensitive to the boson type. The two-body integral is different:
\begin{equation}
\label{norm2b4QED} I_2=4m^2a_1^2{e}_{02}^2\,\tilde{J}_2,
\end{equation}
where
\begin{eqnarray}
\label{norm2b4aQED}
\tilde{J}_2 & = & \frac{1}{4\pi^2}\int_0^{\infty}
dR_{\perp}\,R_{\perp}\int_0^1 dx\,x\nonumber \\
& &\times \sum_{j}
\frac{(-1)^{j}[R_{\perp}^2+(x^2+2x-2)m^2]}{[R_{\perp}^2+(1-x)\mu_j^2+x^2m^2]^2}.\quad
\end{eqnarray}
The integral $\tilde{J}_2$ at $\mu_1\to \infty$ diverges
logarithmically, as $J_2$ in the Yukawa model [see
Eq.~(\ref{J2div})], but with a different coefficient at the
logarithm:
\begin{equation}
\label{J2tQED}
\tilde{J}_2=\frac{1}{8\pi^2}\log\frac{\mu_1}{m}-\frac{1}{4\pi^2}\left(\log\frac{m}{\lambda}
-\frac{9}{8}\right),
\end{equation}
where we assigned a finite (small) mass $\lambda$ to the photon in
order to avoid the infra-red catastrophe. Note that $\tilde{J}_2$
is positive in the limit of infinite PV boson mass.

The normalized solution is given by Eqs.~(\ref{onetwonorm}),
changing $g_{02}$ by $e_{02}$ and $J_2$ by $\tilde{J}_2$.

\subsubsection{Determination of the internal bare coupling constant}
\label{ABCCQED}
Due to the formal coincidence of the state vector structures in
the Yukawa model and QED, we immediately obtain from
Eq.~(\ref{g0b22}):
\begin{equation} \label{g0b22QED}
e_{02}^2=\frac{e^2}{1-e^2\tilde{J_2}}.
\end{equation}

The final solution for the vertex function components
is
\begin{subequations}
\begin{eqnarray}
\label{onebrenormQED}
\psi_1^0=\frac{\sqrt{1-e^2\tilde{J}_2}}{2m}&,&\quad \psi_1^1=0,\\
\label{twobrenormQED}
b_1^{ij}=e&,&\quad b_2^{ij}=0.
\end{eqnarray}
\end{subequations}
It is in full analogy with the corresponding solution for the
state vector in the Yukawa model, given by Eqs.~(\ref{onetworenor}).
%
\subsubsection{Determination of the external bare coupling constant and
calculation of the electromagnetic form factors} \label{ffsQED} 
We
will establish in this section the relationship between the
external electromagnetic BCC and the physical fermion charge and
compute the fermion electromagnetic form factors in the two-body
approximation. The main difference, as compared to the
electromagnetic interaction with a system formed by the Yukawa
interaction, consists in the fact that now the photon emitted (or
absorbed) by the probe is of the same type as photons coming into
the state vector. Formally speaking, the right diagram in
Fig.~\ref{ff2} contains a three-body intermediate state and must
be rejected. However, according to our renormalization scheme
detailed in Sec.~\ref{BCCg}, one should treat the photon
connecting the electromagnetic vertex with the probe as if it was
an external particle, having no relation to the contents of the
state vector. This makes possible the calculation of non-trivial
observables already within the two-body approximation.

In CLFD, the spin-1/2 fermion 3PGF (or electromagnetic vertex,
i.e. the current matrix element between the initial and final
fermion states) has the following general
structure~\cite{km,kms_07} (we have included the physical
electromagnetic coupling constant $e$ into the definition of the
vertex):
\begin{widetext}
\begin{equation}
\label{elmagvert} \bar{u}(p')G^{\rho}u(p)=
e\bar{u}(p')\left[F_1\gamma^{\rho}+
\frac{iF_2}{2m}\sigma^{\rho\nu}q_{\nu}+B_1\left( \frac{{\sla
\omega}}{\omega\cd
p}P^{\rho}-2\gamma^{\rho}\right)+B_2\frac{m{\omega^{\rho}}}{\omega\cd
p}+B_3\frac{m^2{\sla \omega}\omega^{\rho}}{(\omega\cd p)^2}\right]u(p)
\end{equation}
\end{widetext}
with
$\sigma^{\rho\nu}=i(\gamma^{\rho}\gamma^{\nu}-\gamma^{\nu}\gamma^{\rho})/2$
and $P=p'+p$. It is determined by five form factors, the two
physical ($F_{1,2}$) and the three non-physical ($B_{1-3}$) ones.
The non-physical form factors are coefficients at the spin
structures which depend on $\omega$. The term proportional to $B_1$ is constructed in such a way that it gives zero when sandwished between free spinors of momentum $p'=p$. It gives also zero when contracted with $\omega_\rho$. Under the condition
$\omega\cd q=0$ the form factors depend on $Q^2=-q^2$. If the
electromagnetic vertex is calculated exactly or within a given
order of perturbation theory and, moreover, is regularized in a
rotationally invariant way, the non-physical form factors cancel
identically, while the two physical form factors remain, as it
should be. Under approximate non-perturbative calculations, the
non-physical form factors may however survive and plague the
electromagnetic vertex by spurious $\omega$-dependent
contributions. The situation here is fully analogous to that for
the  Yukawa model, where the function $B_{\omega}$ in
Eq.~(\ref{3PGFY}) is just a non-physical scalar form factor. CLFD
allows to separate covariantly the physical and non-physical parts
of the electromagnetic vertex and extract the physical form
factors from the former. In our case, it is enough to contract
both sides of Eq.~(\ref{elmagvert})  with the four-vector
$\omega_{\rho}$:
\begin{equation}
\label{vert+} \bar{u}(p')\omega_{\rho} G^{\rho}u(p)=
e\bar{u}(p')\left[F_1{\sla \omega}+
\frac{iF_2}{2m}\sigma^{\rho\nu}\omega_{\rho}q_{\nu}\right]u(p).
\end{equation}
As we see, the three non-physical form factors $B_{1-3}$
disappeared, since the contraction of  the spin structures proportional to them
with $\omega_{\rho}$ gives zero.
The physical form factors can be found by
the following expressions:
\begin{subequations} \label{F12emv}
\begin{equation}
\label{F1emv} eF_1=\frac{\mbox{Tr}[({\sla
p}'+m)\omega_{\rho}G^{\rho}({\sla p}+m){\sla
\omega}]}{8(\omega\cd p)^2},\\
\end{equation}
\begin{multline}
eF_2=\frac{m}{2(\omega\cd p)Q^2}\\
\label{F2emv}
\times\mbox{Tr}\left[({\sla p}'+m)\omega_{\rho}G^{\rho}({\sla
p}+m)\left(\frac{m{\sla \omega}}{\omega\cd p}-1\right)\right].
\end{multline}
\end{subequations}
Similarly to Eq.~(\ref{vert3PGF}), we write
\begin{equation}
\label{vert12} \omega_{\rho}G^{\rho}\equiv \bar{e}_{02}\bar
G_1+\bar{e}_{01}\bar G_2,
\end{equation}
where
\begin{subequations} \label{onetwoemv}
\begin{equation}
\label{onebemv} \bar G_1=\sum_{i,i'}(-1)^{i+i'}
\psi_1^i\psi_1^{i'}\left[({\sla p}'+m_{i'}){\sla \omega}({\sla
p}+m_i)\right],
\end{equation}
\begin{multline}
\label{twobemv}
\bar G_2=-\frac{1}{(2\pi)^3}\int
d^2R_{\perp}\int_0^1\frac{dx}{2x}\sum_{i,i',j}(-1)^{i+i'+j}\\
\times \frac{\bar{\Gamma}_2^{i'j,\nu}({\sla p}'-{\sla k}_{2j}+m_{i'}){\sla
\omega}({\sla p}-{\sla k}_{2j}+m_{i})\Gamma_2^{ij,\nu}}
{[(p'-k_{2j})^2-m_{i'}^2][(p-k_{2j})^2-m_{i}^2]}.
\end{multline}
\end{subequations}
Again, to make the integral in Eq.~(\ref{twobemv}) convergent, it
is enough to regularize it by the bosonic PV subtraction only. We
can thus retain in the sum the term with $i=i'=0$, neglecting
those with either $i$ or $i'$ equal to 1. According to the
solution~(\ref{twobrenormQED}) for the two-body component of the
state vector,
$\Gamma_2^{ij,\nu}=b_1^{ij}\gamma^{\nu}=e\gamma^{\nu}$. Using the
solution~(\ref{onebrenormQED}) for the one-body component,
substituting Eqs.~(\ref{onetwoemv}) into Eq.~(\ref{vert12}) and
then into each of Eqs.~(\ref{F12emv}), we find
\begin{subequations}\label{F12q}
\begin{eqnarray}
\label{F1q}e F_1&=&\bar{e}_{02}{\cal F}_1^{(1)}+\bar{e}_{01}{\cal F}_1^{(2)}(Q^2),\\
\label{F2q} eF_2&=&\bar{e}_{01}{\cal F}_2^{(2)}(Q^2),
\end{eqnarray}
\end{subequations}
where  $\bar{e}_{01}=e$,
\begin{equation}
\label{F11QED}
{\cal F}_1^{(1)}=1-e^2\tilde{J}_2,
\end{equation}
$\tilde{J}_2$ is given by Eq.~(\ref{norm2b4aQED}), and
\begin{widetext}
\begin{subequations} \label{F12tQED}
\begin{eqnarray}
\label{F12qQED}
{\cal F}_1^{(2)}(Q^2)&=&\frac{e^2}{8\pi^3}\int d^2R_{\perp} \\
&&\int_0^1
dx\sum_{j}
\frac{(-1)^jx[R_{\perp}^2-x{\bf R}_{\perp}{\bg
\Delta}+(x^2+2x-2)m^2-(1-x){\Delta}^2]}{[R_{\perp}^2+(1-x)\mu_j^2+x^2m^2] [({\bf
R}_{\perp}-x{\bg \Delta})^2+(1-x)\mu_j^2+x^2m^2]},\nonumber\\
\label{F22qQED} {\cal F}_2^{(2)}(Q^2)&=&\frac{e^2m^2}{4\pi^3}\int d^2R_{\perp} \\
&&\int_0^1
dx\sum_{j}
\frac{(-1)^j x^2(1-x)}{[R_{\perp}^2+(1-x)\mu_j^2+x^2m^2]
[({\bf R}_{\perp}-x{\bg \Delta})^2+(1-x)\mu_j^2+x^2m^2]}.\nonumber
\end{eqnarray}
\end{subequations}
\end{widetext}
Here $\mu_0\equiv \mu=0$. Note that ${\cal F}_1^{(2)}(0)=e^2\tilde{J}_2$
which is just the norm of the two-body sector.

The requirement $F_1=1$ at $Q^2=0$ leads to the relation
\begin{equation}
eF_1(Q^2=0)\equiv e=\bar{e}_{02}(1-e^2\tilde{J}_2)+e^3\tilde{J}_2,
\end{equation}
so that
\begin{equation}
\label{e02QED} \bar{e}_{02}=e.
\end{equation}
Hence, the electromagnetic external BCC calculated in the two-body
approximation is not renormalized. Such a coincidence happens not
by chance, but reflects a general property of the theory. We will
discuss this fact in more detail below.

From Eqs~(\ref{F12q}) we find the renormalized form factors:
\begin{subequations}
\begin{eqnarray}
\label{F1qpertQED}
F_1&=&1+{\cal F}_1^{(2)}(Q^2)-{\cal F}_1^{(2)}(0),\\
\label{F2qpertQED}
F_2&=&{\cal F}_2^{(2)}(Q^2).
\end{eqnarray}
\end{subequations}
This result exactly coincides with that found in the second order
of perturbation theory, though no expansions in the coupling
constant have been done. A similar issue was obtained above for the
Yukawa model [see Eq.~(\ref{FYUK})].

Expressing $e^2$  through the fine structure constant $\alpha$ by
the relation $e^2=4\pi\alpha$ and calculating the
integrals~(\ref{F12tQED}), we obtain (for simplicity, we
decomposed the form factors in powers of $Q^2$, up to second
order):
\begin{subequations} \label{F12fin}
\begin{eqnarray}
\label{F1finQED}
F_1&=&1-\frac{\alpha Q^2}{3\pi m^2}
\left(\log\frac{m}{\lambda}-\frac{3}{8}\right)+O(Q^4),\quad\\
\label{F2finQED}
F_2&=&\frac{\alpha}{2\pi}-\frac{\alpha Q^2}{12\pi m^2}+O(Q^4).
\end{eqnarray}
\end{subequations}
The formulas~(\ref{F12fin}) exactly coincide with the familiar
perturbative expressions. In particular, $F_2$ at $Q^2=0$
reproduces the well-known Schwinger correction~\cite{sch} to the
electron anomalous magnetic moment. As expected, the form factors
have finite limits, when the PV masses tend to infinity.

The external and internal BCC's calculated in the two-body
approximation and given, respectively, by Eqs.~(\ref{e02QED})
and~(\ref{g0b22QED}) do not coincide, as in the Yukawa model. This
fact has been already explained in the end of Sec.~\ref{ffsYuk}.
In QED however, we encounter a new effect, namely, the identity
between the external BCC $\bar{e}_{02}$ and the physical electron
charge $e$.
 It is a specific feature of QED, which appears as a consequence of the Ward identity.
 To illustrate the situation,
 one may come back to Fig.~\ref{RC2}.  Due
 to the Ward identity,
the total contribution of the perturbative diagrams (a)--(e) gives
zero, and the external BCC ${\bar e}_{02}$ is not renormalized at all.
The same arguments hold true for the non-perturbative calculations
in two-body truncated Fock space. As we will see in the next
section, this takes place for truncations of any order.

One more point is also worth mentioning. Neither the perturbative
expansion of the internal BCC, Eq.~(\ref{g0b22QED}), at
$\mu_1\to\infty$,
$$
e_{02}^2=e^2\left(1+\frac{\alpha}{4\pi}\log\frac{\mu_1^2}{m^2}\right),
$$
nor the external BCC $\bar{e}_{02}^2=e^2$ coincide with the
well-known purely perturbative formula relating the BCC with the
physical charge and given in most textbooks on quantum field
theory:
\begin{equation}
\label{pert}
 e_{0,pert}^2=e^2\left(1+\frac{\alpha}{3\pi}\log\frac{\Lambda^2}{m^2}\right),
\end{equation}
where $\Lambda^2$ is an invariant cutoff which can be identified,
in the logarithmic approximation, with $\mu_1^2$. This result is
nevertheless quite natural, since Eq.~(\ref{pert}), besides the
contributions shown in Fig.~\ref{RC2}(a-e) (which, being summed
up, give zero due to the Ward identity), takes into account also
the effect of vacuum polarization, i.e. the diagram with the
electron-positron loop on the external photon line, shown in
Fig.~\ref{RC2}(f). Evidently, our two-body approximation is unable
to embrace such an effect. The latter can be incorporated when
Fock sectors containing electron-positron pairs are included into
the state vector.

\subsubsection{External bare coupling constant. What happens for $N >2$ ?}
\label{N3}
\begin{figure}[btph]
\begin{center}
\includegraphics[width=20pc]{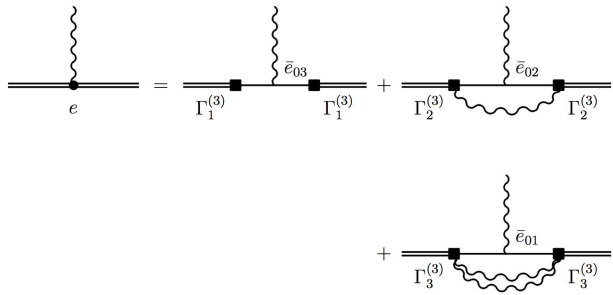}
\caption{Electron electromagnetic vertex for the three-body Fock
space truncation. \label{WI3}}
\end{center}
\end{figure}

We can easily generalize the result~(\ref{e02QED}) to the case
where an arbitrary number of photons in the state vector is
considered. We can do that by recurrence. In the $N=3$ case
(one electron and no more than two photons), for instance, the
form factors are given by the contributions  from the diagrams
shown in  Fig.~\ref{WI3}.
At $Q^2=0$ we have
\begin{equation}
e=\bar{e}_{03} {\cal F}_1^{(1)}+\bar{e}_{02} {\cal
F}_1^{(2)}(Q^2=0) +\bar{e}_{01} {\cal F}_1^{(3)}(Q^2=0).
\end{equation}
The functions ${\cal F}_1^{(1,2,3)}$ are just the contributions of
the amplitudes of the three diagrams on the r.-h.s. of the
graphical equation shown in Fig.~\ref{WI3}. For QED, one can
easily check that ${\cal F}_1^{(n)}(Q^2=0)=I_n$,  where $I_n$ is
the contribution of the $n$-body Fock sector to the norm of the
state vector. With $\bar{e}_{01}=\bar{e}_{02}=e$ we get
\begin{equation}
e=\bar{e}_{03} I_1+e(I_2+I_3).
\end{equation}
Due to the normalization condition, $I_1+I_2+I_3=1$. We thus have
\begin{equation}
e=\bar{e}_{03}\left[1-(I_2+I_2)\right]+e(I_2+I_3),
\end{equation}
so that
\begin{equation} \label{WIn}
\bar{e}_{03}=e.
\end{equation}
Repeating the same arguments to the cases of $N=4,\,5,\ldots$ Fock
space truncations, we conclude that Eq.~(\ref{WIn}) holds to all
orders in $N$. Hence,
 our formalism  respects the requirement of the Ward
identity even in truncated Fock space.\\

\section{Application to the N=3 Fock space truncation in a scalar
model} \label{scalar} 
We re-examine here our previous calculation
\cite{bckm} of the state vector in a pure scalar model for the
three-body Fock space truncation, using the renormalization
strategy outlined above. We shall see how this strategy should be
applied in the context of a non-perturbative physical problem
where the results  can not be reproduced in
perturbation theory, unlike the cases of the two-body
approximation in the Yukawa model and QED.

 Though all the particles we consider are spinless, we will distinguish two
 types of them, described by
the free fields $\varphi(x)$ and $\chi(x)$ and related to the
"main" (i.e. analogous to fermions in the Yukawa model or QED) and
"exchanged" particles. For shortness, we will refer to them
 as scalar "nucleons" and bosons, respectively. The interaction Hamiltonian is
 \begin{equation}
 \label{Hamscal}
 H^{int}(x)=-g_0\varphi^2\chi'-\delta m^2\varphi^2,
 \end{equation}
  where the prime at $\chi$ denotes the fact that it is a sum of a physical and a PV component.
  Since all divergencies
 are regularized already by the PV boson, we do not need to introduce a PV "nucleon".
 The Hamiltonian~(\ref{Hamscal})
 contains the BCC $g_0$ and the MC $\delta m^2$ which must be adjusted to reproduce
 correctly the physical "nucleon" mass $m$ and the physical coupling constant $g$.
 
\begin{figure}[btph]
\begin{center}
\includegraphics[width=21pc]{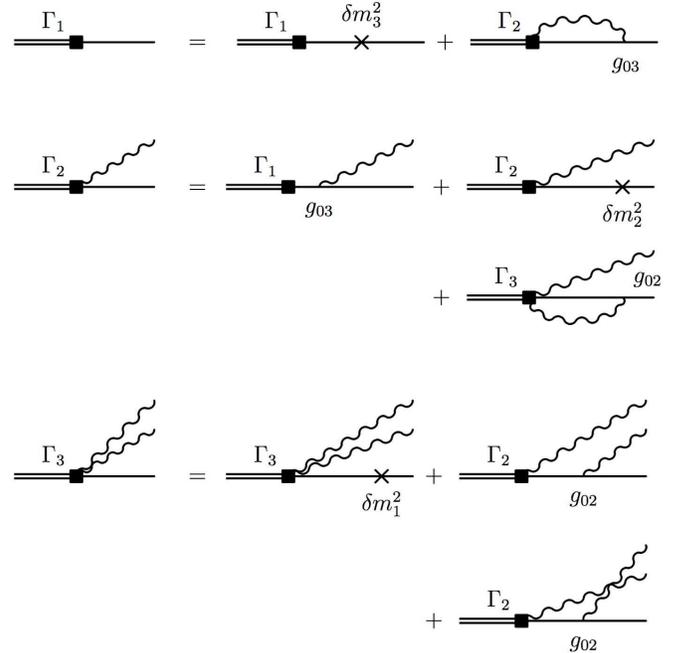}
\caption{System of equations for the Fock components in a
three-body pure scalar model. \label{scalarfig}}
\end{center}
\end{figure}

 The system of eigenstate equations for the vertex functions in the three-body approximation
is shown graphically in Fig.~\ref{scalarfig}.  Solid and wavy lines correspond
to "nucleons" and bosons.
In analytical form the system of equations reads
\begin{widetext}
\begin{subequations} \label{eq11}
\begin{eqnarray}
\Gamma_1&=&\frac{\delta m_{3}^{2}\Gamma_1}{m^{2}-M^{2}}+
g_{03}\sum_{j}\frac{(-1)^j}{(2\pi)^3}\int d^{2}R_{\perp}
\int_0^1\frac{dx}{2x(1-x)}\,
\frac{\Gamma_{2}^j(R_{\perp}, x)}{s_{12}^{j}-M^2}, \label{eq11.a}\\
\Gamma_2^j(R_{\perp}, x) &=&\frac{g_{03}\Gamma_1}{m^{2}-M^{2}}
+\frac{\delta m_{2}^{2}\Gamma_2^j(R_{\perp}, x) }{(1-x)(s_{12}^j-M^{2})} \nonumber\\
&&+g_{02}\sum_{j'}\frac{(-1)^{j'}}{(2\pi)^3} \int
d^{2}R'_{\perp}\int_0^{1-x}dx' \frac{\Gamma_{3}^{jj'}({
R}_{\perp}, x; { R}'_{\perp},
x')}{2x'(1-x-x')(s_{123}^{jj'}-M^2)},
\label{eq11.b}\\
\Gamma_3^{jj'}(R_{\perp}, x; R'_{\perp}, x')&=&
\frac{g_{02}\Gamma_2^j(R_{\perp }, x)}{(1-x)(s_{12}^j-M^2)}+
\frac{g_{02}\Gamma_2^{j'}(R'_{\perp },
x')}{(1-x')(s_{13}^{j'}-M^2)} \label{eq11.c}
\end{eqnarray}
\end{subequations}
\end{widetext}
with
\begin{eqnarray*}
s_{12}^j&=&\frac{R_{\perp}^{2} + m^2}{(1-x)} +
\frac{R_{\perp}^{ 2} + \mu^2_j}{x}, \\
s_{13}^{j'}&=&\frac{{R'}_{\perp}^{2} + m^2}{(1-x')} +
\frac{{R'}_{\perp}^{ 2} + \mu^2_{j'}}{x'},\\
s_{123}^{jj'}&=&\frac{({\bf R}_{\perp} + {\bf R}'_{\perp})^{2} + m^2}{(1-x-x')} +
\frac{R_{\perp}^{ 2} + \mu^2_j}{x}
+ \frac{{R'}_{\perp}^{ 2} + \mu^2_{j'}}{x'}.
\end{eqnarray*}
In deriving Eqs.~(\ref{eq11}) we took into account that $\delta
m_1^2=0$, since no "nucleon" mass renormalization occurs for the
one-body Fock space truncation. As before, the index $j=0$ relates
to the physical boson with the mass $\mu_0\equiv \mu$, while $j=1$
--- to the PV one with the mass $\mu_1$. We also omitted, for
simplicity, the superscript $(3)$ at each vertex function.
Necessary kinematical relations needed to obtain Eqs.~(\ref{eq11})
are presented in Appendix~\ref{kin3}.

According to our strategy,  the internal BCC $g_{02}$ and the MC
$\delta m^2_{2}$  should be determined from the two-body
approximation. Using the results of Ref.~\cite{bckm}, we have
\begin{subequations} \label{bgmd}
\begin{eqnarray}
g_{02}&=&\frac{g}{\sqrt{1-g^2 \bar I_2} },
\label{bg02} \\
\delta m^2_{2} &=& -g_{02}^2 \bar \Sigma(p^2=M^2) \label{dm02},
\end{eqnarray}
\end{subequations}
with
$$
\bar I_2=\sum_j \frac{(-1)^j}{16\pi^3}
\int d^2R_{\perp} \int_0^1 \frac{dx}{x(1-x)(s_{12}^j-M^2)^2}
$$
and
$$
\bar \Sigma(p^2)=\sum_j \frac{(-1)^j}{16\pi^3} \int d^2R_{\perp}
\int_0^1 \frac{dx}{x(1-x)(s_{12}^j-p^2)},
$$
being, respectively, the norm of the two-body sector and the
"nucleon" self-energy amputated from the coupling constant
squared. In the limit $\mu_1\to \infty$, the two-body norm tends
to a finite limit, while the self-energy diverges logarithmically.

The three-body component $\Gamma_3$ is expressed through the
two-body one, $\Gamma_2$, and can be eliminated from the system of
equations~(\ref{eq11}). Since the latter is homogeneous, we can
start with a value of $\phi_1\equiv\Gamma_1/(m^2-M^2)$ equal to
one. Dividing Eqs.~(\ref{eq11.b}) and~(\ref{eq11.c}) by $g_{03}$,
we get a system of  inhomogeneous equations for the
(non-normalized) state vector characterized by the Fock components
$\bar \Gamma_{2,3}\equiv \Gamma_{2,3}/g_{03} $. Excluding
$\Gamma_1$ and $\Gamma_3$ from Eqs.~(\ref{eq11}) and going over to
the limit $M\to m$,  we obtain a closed inhomogeneous equation
involving the two-body component only:
\begin{multline}
\label{gam2}
\bar \Gamma_2^j(R_{\perp},x)=1+ \frac{g_{02}^{2}\left[\bar
\Sigma(s^j_{1})-\bar\Sigma(m^2)\right]} {m^2-s_1^j}\bar
\Gamma_2^j(R_{\perp},x) \\
+\frac{g_{02}^{2}}{16\pi^3}\sum_{j'}(-1)^{j'}\int
d^2R'_{\perp}\int_0^{1-x}\frac{dx'}{x'(1-x')}\\
\times\frac{\bar \Gamma_{2}^{j'}(R'_{\perp},x')}{(1-x-x')
(s_{13}^{j'}-m^2)(s_{123}^{jj'}-m^2)},
\end{multline}
where
\begin{equation}
\label{s1j}
s_1^j=-\frac{R^2_{\perp}}{x}+(1-x)m^2-\frac{1-x}{x}\mu_j^2.
\end{equation}

The solution $\bar \Gamma_{2}^j$ of Eq.~(\ref{gam2}) can be found
numerically for given values of $m$ and $g$. It is completely
finite in the limit $\mu_1\to\infty$. To get the normalized vertex
functions, we should calculate the norm $N^{(3)}$ of the state
vector.  It can be written schematically as
\begin{equation}
\label{norm3} N^{(3)}\equiv N_1+N_2+N_3=1+g_{03} ^2(\bar N_2 +
\bar N_3) ,
\end{equation}
where $\bar N_{2,3}$ are the two- and three-body norms written in
terms of $\bar \Gamma_{2,3}$.
The normalized components $\Gamma_{1,2,3}^R$ are thus given by

\begin{equation}
\label {renorwf}
 \Gamma_{1,2,3}^R=\frac{\Gamma_{1,2,3} }{\sqrt{N^{(3)}}} .
 \end{equation}

We still have to fix the internal BCC $g_{03}$ to calculate the
normalized vertex functions. It is found from the requirement that
the value of the physical (i.e. taken for $j=0$) two-body vertex
function $\Gamma_2^R$ in the kinematical point ($s^*\equiv
s_{12}^{j=0}=m^2, x^*)$, where $x^*$ is a given fixed value of $x$, equals the physical coupling constant:
\begin{equation}
\label{physcond}
\Gamma_2^{R}(s^*, x^*) = g.
\end{equation}
 Let us call $\bar \Gamma_2^{*}$ the value of $\bar \Gamma_2^{R}$ in this point.
 From Eqs.~(\ref{norm3})--(\ref{physcond}) we get
\begin{equation}Ê
\label{ggob} \frac{g_{03}}{ \sqrt{1+g_{03}^2(\bar N_2 + \bar
N_3)}} \bar \Gamma_2^{*} = g,
\end{equation}
which gives
\begin{equation}\label{g03}
g_{03}=\frac{g}{\sqrt{ \left[\bar \Gamma_2^{*}\right]^2 -
g^2(\bar N_2 + \bar N_3)}}.
\end{equation}
We can see that Eq.~(\ref{g03}) is just a generalization of
Eq.~(\ref{bg02}), since in the absence of the three-body component
we have $\bar  \Gamma_2^{*} = 1, \bar N_3 = 0$, and $\bar N_2 =
\bar I_2$. Using Eqs.~(\ref{renorwf}) and~(\ref{ggob}), we can
represent the normalized vertex function $\Gamma_{2}^R$ simply as
\begin{equation}
\label{G2R}
\Gamma_2^R=g\frac{\bar\Gamma_2}{\bar \Gamma_{2}^{*}},
\end{equation}
which allows to find it directly from the solution of the
inhomogeneous equation~(\ref{gam2}), without calculating
explicitly the norm $N^{(3)}$. The latter is just needed to
calculate the normalized one-body component $\phi_1^R =
1/\sqrt{N^{(3)}}$.

The substitution of  Eq.~(\ref{g03})  into Eq.~(\ref{norm3})
yields
\begin{equation}
N^{(3)}=\frac{\left[\bar \Gamma_{2}^{*}\right]^2}
{\left[\bar\Gamma_{2}^{*}\right]^2-g^2(\bar N_2 + \bar N_3)}.
\end{equation}
Finally, substituting the normalized functions
$\Gamma_1^R=\phi_1^R(m^2-M^2)$ and $\Gamma_2^R$ into
Eq.~(\ref{eq11.a}), we arrive, in the limit $M\to m$, at the
following expression for the MC:
\begin{eqnarray}
\delta m_3^2&=&-\frac{g^2}{\left[\bar\Gamma_{2}^{(*)}\right]^2-g^2(\bar N_2 + \bar N_3)}
\sum_{j}\frac{(-1)^j}{(2\pi)^3}\nonumber \\
&&\times\int d^{2}R_{\perp} \int_0^1\frac{dx}{2x(1-x)}\,
\frac{\bar\Gamma_{2}^j(R_{\perp}, x)}{s_{12}^{j}-m^2}.\quad\quad
\end{eqnarray}
The normalized three-body vertex function is found from
Eq.~(\ref{eq11.c}), changing $\Gamma_2$ to $\Gamma_2^R$. The norms
$\bar N_n$ are calculated according to
\begin{eqnarray}
\bar N_n&=&\frac{2}{(2\pi)^{3(n-1)}(n-1)!}\nonumber \\
&&\times\int \prod_{l=1}^n
\frac{d^2R_{l\perp}dx_l}{2x_l}\left[\frac{\bar \Gamma_n^2}{(s_n-m^2)^2}\right]\nonumber \\
&&\times\delta^{(2)}\left(\sum_{l=1}^n {\bf R}_{l\perp}\right)
\delta\left(\sum_{l=1}^n x_n-1\right), \label{N2bN3b}
\end{eqnarray}
where $s_n=\sum_{l=1}^n(R_{l\perp}^2+m_l^2)/x_l$ and, in the case
of the PV regularization,
$$
\bar \Gamma_n^2=\sum_{j_1\ldots j_{n-1}}(-1)^{j_1+\ldots +j_{n-1}}\bar
\Gamma_n^{j_1\ldots j_{n-1}}
\bar \Gamma_n^{j_1\ldots j_{n-1}}.
$$
In such a way, all unknown parameters are expressed through the
solution of the inhomogeneous equation~(\ref{gam2}).

The application of this strategy to the three-body Fock space
truncation in the Yukawa model (one fermion and two scalar bosons)
is in progress.

\section{Concluding remarks and perspectives}
\label{conc} We have presented a systematic strategy to calculate
physical observables in CLFD, when Fock space is truncated. This
requires to implement an appropriate renormalization scheme in a
non-perturbative framework. Within CLFD, we have shown how to fix
the counterterms and the BCC's
 of the elementary Hamiltonian in a consistent
way. As a check of our formalism, we treated  the Yukawa model and
QED in the two-body approximation, as well as a pure scalar model
in the three-body approximation. We were able to recover, for the
first time, the standard renormalization of the electromagnetic
charge according to the Ward identity, without any perturbative
expansion. This shows that no divergences are left uncancelled in
the above calculations. Moreover, the first correction to the
electron anomalous magnetic moment (the Schwinger correction) is
recovered analytically.

Our results have been made possible because of the following three
important features of the formalism:\\

{(i)} First we can extract the physical part of the two-body
vertex function in the point $s=m^2$, in order to identify it with
the physical coupling constant. This part is explicit in our
formalism, since it should be independent of the LF plane
orientation  determined by the four-vector  $\omega$.\\

{(ii)} The counterterms and the BCC do depend on the Fock sector,
in order to cancel all divergences. We give a well-defined
systematic procedure to calculate them.\\

{(iii)} We have to distinguish two types of BCC's: those of the
first type are used to calculate the state vector itself, by means
of the eigenstate equation, while the BCC's of the second type are
responsible for describing interactions of the constituents with
external particles (electromagnetic probes, for example). As we
increase the number of Fock sectors, the two BCC's of different
types are expected to converge to the same limit
(for finite PV masses).\\

Our results are very encouraging in the perspective of doing true
non-perturbative calculations of bound state systems in a field
theoretical framework. In case of success, they may become a real
alternative for lattice calculations.

\begin{acknowledgments}
Two of us (V.A.K. and A.V.S.) are sincerely grateful for the warm
hospitality of the Laboratoire de Physique Corpusculaire,
Universit\'e Blaise Pascal, in Clermont-Ferrand, where part of
the present study was performed. This work has been partially
supported by the RFBR grant No. 05-02-17482-a.
\end{acknowledgments}

\appendix

\section{Light-front Hamiltonian in the Yukawa model with the Pauli-Villars regularization}
\label{appA}
We shall find in this section the interaction LF Hamiltonian
incorporating PV fields.  The methodology is the same as in
Ref.~\cite{kms_04}. For this reason, we will not give here the
detailed algebra with all explicit intermediate results, but outline
only the main steps of the procedure.

First, substituting the
Lagrangian~(\ref{lagrPVfull}) into Eq.~(\ref{emt}) and keeping in
mind that now $Y_i$ runs through the whole set of the physical and
PV fields, we construct the corresponding energy-momentum tensor
and then find the four-momentum operator in terms of the
Heisenberg fields. In order to have the interaction part of
the four-momentum operator in Schr\"{o}dinger representation, we
should find the constraints which connect different field
components at the same time and distinguish the independent
components. We then express the four-momentum operator through the
independent field components only and change after that all the
Heisenberg operators by the free ones.

It is convenient to perform calculations in the reference frame
where the four-vector $\omega^{\rho}$ has the components
$(1,0,0,-1)$ [the four-vector $\omega_{\rho}$ is
(1,0,0,1)].\footnote{It is not necessary to fix any particular
reference frame. All subsequent algebraic manipulations, in
principle, can be done in explicitly covariant notations, however,
the calculations would be more lengthy.} Under this condition the
LF "time" is $x^+=x^0+x^3$, while the "coordinates" are
$x^-=x^0-x^3$ and ${\bf x}^{\perp}=(x^1,x^2)$. Analogously, we
introduce the plus-, minus-, and transverse components for any
four-vector.

The equations of motion for the fermionic fields are
\begin{subequations} \label{eqmottot}
\begin{eqnarray}
\label{eqmotpsi}
(i\gamma^{\nu}\partial_{\nu}-m)\Psi & = & -(g_{0}\Phi'+\delta m)\Psi',\\
\label{eqmotpsiPV}
(i\gamma^{\nu}\partial_{\nu}-m_1)\Psi_{PV} & = & (g_{0}\Phi'+\delta m)\Psi',
\end{eqnarray}
\end{subequations}
Splitting the bispinors into the so-called plus- and
minus-components (they should not be confused with the plus- and
minus-components of four vectors)
\begin{equation}
\Psi=\Psi^{(+)}+\Psi^{(-)},
\end{equation}
where
\begin{equation}
\label{psipm}
\Psi^{(\pm)}=\Lambda^{(\pm)}\Psi,\,\,\,\,\,\,\,\,
\Lambda^{(\pm)}=\frac{1}{2} (1\pm \gamma^0\gamma^3)
\end{equation}
(and similarly for $\Psi_{PV}$), we can express, by means
of Eqs.~(\ref{eqmottot}), the
minus-components through the plus-ones:
\begin{subequations} \label{constr12}
\begin{eqnarray}
\label{constr1} \Psi^{(-)} & = &
\frac{\gamma^0}{i\partial^+}\left[(i{\bf \partial}^{\perp}{ \bg
\gamma}^{\perp}+m)\Psi^{(+)}\right.\nonumber \\
&&\left.-(g_{0}\Phi'+\delta m){\Psi'}^{(+)}\right],\\
\label{constr2} \Psi^{(-)}_{PV} & = &
\frac{\gamma^0}{i\partial^+}\left[(i{\bf \partial}^{\perp}{ \bg
\gamma}^{\perp}+m_1) \Psi^{(+)}_{PV}\right.\nonumber \\
&&\left.+ (g_{0}\Phi'+\delta m){\Psi'}^{(+)}\right].
\end{eqnarray}
\end{subequations}
The operator $1/(i\partial^+)$ acting on an arbitrary
function $f$ of the coordinate $x$ involves integration over the
minus-component of its argument:
\begin{equation}
\label{1/d}
\frac{1}{i\partial^+}f(x^-)=
-\frac{i}{4}\int_{-\infty}^{+\infty}dy^-\,\epsilon(x^--y^-)f(y^-),
\end{equation}
where $\epsilon$ is the sign function.

Eqs.~(\ref{constr12}) are constraints, since they do not include
the "time" derivatives $\partial^{-}$ and connect field components
at any time. We see that the minus-components of the spinor fields
are not independent in the sense that they are expressed through
$\Psi^{(+)}$, $\Psi^{(+)}_{PV}$, $\Phi$, and $\Phi_{PV}$. Taking
the four latter quantities as independent fields, we can express
the four-momentum operator $\hat{P}_{\rho}$ through them.
After that, finding this operator in Schr\"{o}dinger or
interaction representation reduces to simple changing the
Heisenberg fields $\Psi^{(+)}$, $\Psi^{(+)}_{PV}$, $\Phi$, and
$\Phi_{PV}$ by their free counterparts $\psi^{(+)}$,
$\psi^{(+)}_{PV}$, $\varphi$, and $\varphi_{PV}$. Splitting, as in
Eq.~(\ref{pshr}), the four-momentum operator into the free and
interaction parts, and keeping in mind that
$\omega_+=(\omega_0+\omega_3)/2=1$, ${\bg \omega}_{\perp}=-{\bg
\omega}^{\perp}={\bf 0}$, $\omega_-=(\omega_0-\omega_3)/2=0$, we
obtain from Eq.~(\ref{pintham}):
\begin{equation}
\label{Pham} \hat{P}^{int}_+=\frac{1}{2}\int
d^2x^{\perp}dx^-\,H^{int}_{PV}(x).
\end{equation}
Calculations performed according to the rules given above yield
the formula~(\ref{hamPV}) for the interaction Hamiltonian.

\section{CLFD graph technique rules}\label{appB}
The graph technique rules for CLFD have already been given in
Ref.~\cite{cdkm} for the calculation of the $S$-matrix in terms of
the operator $-\tilde{H}_{int}$. Since the eigenstate equation
(\ref{eq3}) is also expressed in terms of the same quantity,
these rules are applicable in our case too.

\begin{figure}[btph]
\begin{center}
\includegraphics[width=20pc]{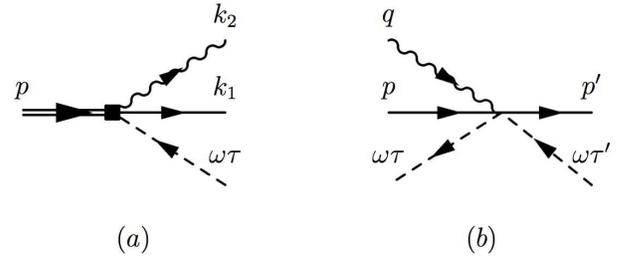}
\caption{Convenient representation of the vertex function (a) and
a typical elementary vertex (b) in terms of the spurion line.
\label{spurion}}
\end{center}
\end{figure}

Because of the particular conservation law~(\ref{k1n}), it is
convenient to represent, for the calculation of CLFD diagrams, 
the momentum $\omega
\tau_n$ by a separate line, in addition to the ordinary particle
lines. This is the so-called spurion line~\cite{cdkm}. For
practical calculations, one may thus represent the vertex
function, as well as any elementary vertex, by the diagrams
indicated in Fig.~\ref{spurion}. The spurions are shown by the
dashed lines.
The momentum conservation laws associated to the vertices (a) and
(b) in Fig.~\ref{spurion} are, respectively
$$
\delta^{(4)}(p+\omega \tau -k_1 - k_2)
$$
and
$$
\delta^{(4)}[p+q-p'+\omega (\tau ' - \tau)].
$$
We emphasize that spurions are not considered as true particles. 
They serve simply for a convenient representation of the departure
of intermediate particles from the energy shell.

Appearance of PV fields changes the CLFD graph technique rules, as
compared to those listed in Ref.~\cite{cdkm}. However, one does not need 
to re-derive the rules from the
very beginning  in order to
incorporate the changes. Indeed, PV particles can be considered as new
fermions and bosons having their own masses and negative norms.
Hence, they participate in the construction of CLFD diagrams on equal
grounds with physical particles, simply increasing the number of
diagrams. The analytical rules to calculate amplitudes of the
latter remain almost the same, excepting the expressions for
particle propagators. Since physical and PV fields represent two
different groups of particles, they can not pair between
themselves, i.e. any line in the diagrams relates to either a
physical or a PV particle. Pairings within each group are allowed.
PV propagators correspond to the PV particle masses and also
differ by a sign from the physical ones. In momentum space the
propagators are
$$
(-1)^{j}\theta(\omega\cd p)\delta(p^2-\mu_j^2)
$$
for bosons and
$$
(-1)^{i}({\sla p}+m_i)\theta(\omega\cd p)\delta(p^2-m_i^2)
$$
for fermions. The indices $i$ and $j$ describe the particle type:
$i=0$ and $j=0$ correspond to physical particles, while $i=1$
and $j=1$ relate to PV ones.

\section{Kinematical relations}
\label{kin}
\subsection{One- and two-body kinematics}
\label{kin1} To calculate the amplitudes of the CLFD diagrams
contributing to Eqs.~(\ref{eq2t}) and~(\ref{V22t}), we need to
express the intermediate momenta and their contractions with the
external momenta through the variables ${\bf R}_{\perp}$, ${\bf
R}'_{\perp}$, $x$, and $x'$.

The momentum conservation laws in the vertices  lead to the following
equalities:
\begin{subequations}
\begin{eqnarray}
\label{Ap1}
p_{1i}&=&p+\omega\tau_{1},\label{Ap1a}\\
p'_{1i'}&=&p+\omega\tau'_{1},\label{Ap1b}\\
k_{1i}&=&p+\omega\tau_{2}-k_{2j},\label{Ap1c}\\
k'_{1i'}&=&p+\omega\tau'_{2}-k'_{2j'}.\label{Ap1d}
\end{eqnarray}
\end{subequations}
Squaring the first two of these equations and taking into account
that $p^2=M^2$, $p_{1i}^2=m_{i}^2$, ${p'}_{1i'}^2=m_{i'}^2$, and
$\omega^2=0$, we get
\begin{equation}
\label{Ap2} \tau_{1}=\frac{m_{i}^2-M^2}{2\omega\cd
p},\quad \tau'_{1}=\frac{m_{i'}^2-M^2}{2\omega\cd p}.
\end{equation}
Since, by the definition~(\ref{lf}), $\omega\cd (k_{2j}-xp)=0$, we
have $(k_{2j}-xp)^2= -({\bf k}_{2j\perp}-x{\bf
p}_{\perp})^2=-{R}_{\perp}^2$. This allows to represent the scalar
product $p\cd k_{2j}$ as
\begin{equation}
\label{Ap4}
p\cd k_{2j}=\frac{{R}_{\perp}^2+M^2{x}^2+\mu_{j}^2}{2x}.
\end{equation}
For the scalar product $p\cd k'_{2j'}$ we analogously find
\begin{equation}
\label{Ap4a}
p\cd k'_{2j'}=\frac{{R'}_{\perp}^2+M^2{x'}^2+\mu_{j'}^2}{2x'}.
\end{equation}
Squaring Eq.~(\ref{Ap1c}) under the condition
${k}_{1i}^2=m_{i}^2$, we obtain with the help of Eq.~(\ref{Ap4}):
\begin{eqnarray}
\label{Ap5}
\tau_{2}&=&\frac{m_{i}^2-(p-k_{2j})^2}{2\omega\cd (p-k_{2j})}\nonumber \\
&=&\frac{1}{2\omega\cd p}\left[
\frac{{R}_{\perp}^2+\mu_{j}^2}{x}+\frac{{R}_{\perp}^2+m_{i}^2}{1-x}-M^2\right].\quad
\end{eqnarray}
Analogously,
\begin{equation}
\label{Ap5a}
\tau'_{2}=\frac{1}{2\omega\cd p}\left[
\frac{{R'}_{\perp}^2+\mu_{j'}^2}{x'}+\frac{{R'}_{\perp}^2+m_{i'}^2}{1-x'}-M^2\right].
\end{equation}
On the other hand,  since $(p+\omega\tau'_2)^2=(k'_{1i'}+k'_{2j'})^2\equiv s_{12}^{i'j'}$,
we have
\begin{eqnarray}
\label{s12ipjp}
s_{12}^{i'j'}&=&M^2+2(\omega\cd p)\tau'_2\nonumber \\
&=&\frac{{R'}_{\perp}^2+\mu_{j'}^2}{x'}+\frac{{R'}_{\perp}^2+m_{i'}^2}{1-x'}.
\end{eqnarray}
The above equalities are enough to calculate the
self-energy~(\ref{sigma}) and to obtain the system of equations
(\ref{ur2b}) for the scalar components of the vertex functions
from the initial system of equations~(\ref{eq2t}).

\subsection{Two-body contribution to the three-point Green's function}
\label{kin2}

To calculate the two-body contribution to the scalar and
electromagnetic form factors from the vertices~(\ref{twob3PGF})
and~(\ref{twobemv}), we need to express the contractions of the
four-vector $k_{2j}$ with the external momenta $p$ and $p'$
through the integration variables ${\bf R}_{\perp}$ and $x$. The
contraction $p\cd k_{2j}$ is given by Eq.~(\ref{Ap4}). In order to
find the scalar product $p'\cd k_{2j}$, we introduce the
four-vector $R'\equiv k_{2j}-xp'$ and consider its properties. As
$\omega\cd p=\omega\cd p'$, we have $\omega\cd R'=0$ and, hence,
$$
{R'}^2=-{R'}_{\perp}^2=-({\bf k}_{2j\perp}-x{\bf p}_{\perp}-x{\bf q}_{\perp})^2
$$
\begin{equation}
\label{Ap6}
=({\bf R}_{\perp}-x{\bg \Delta})^2.
\end{equation}
On the other hand,
\begin{equation}
\label{Ap7}
{R'}^2=(k_{2j}-xp')^2=\mu_j^2-2x(p'\cd k_{2j})+x^2m^2.
\end{equation}
Comparing the r.-h.s.'s of Eqs.~(\ref{Ap6}) and~(\ref{Ap7}), we
finally get
\begin{equation}
\label{Ap8}
p'\cd k_{2j}=\frac{({\bf R}_{\perp}-x{\bg \Delta})^2+\mu_j^2+x^2m^2}{2x}.
\end{equation}

\subsection{Three-body kinematics}
\label{kin3} To cast the equations for the vertex functions in the
three-body approximation to the form~(\ref{eq11}), we should
define due kinematical variables. For this purpose, we consider a
set of three four-vectors
\begin{equation}
\label{Ap9}
R_n=k_n-x_np,
\end{equation}
where $n=1,\,2,\,3$ and $x_n=(\omega\cd k_n)/(\omega\cd p)$. The four-vectors
$k_n$ are just the four-momenta of the constituents. For shortness, we do not pay attention
to the particle type (physical or PV one). The momentum conservation
law reads
\begin{equation}
\label{Ap10}
\sum_n k_n=p+\omega\tau_3,
\end{equation}
 where $\omega\tau_3$ is the spurion momentum in the three-body state.
Contracting both sides of Eq.~(\ref{Ap10}) with $\omega$, we have
\begin{equation}
\label{Ap11}
\sum_n x_n=1.
\end{equation}
The substitution of $k_n=R_n+x_np$ into Eq.~(\ref{Ap10}) gives
\begin{equation}
\label{Ap12} \sum_n R_n=\omega\tau_3
\end{equation}
and, hence,
\begin{equation}
\label{Ap13}
\sum_n {\bf R}_{n\perp}={\bf 0}.
\end{equation}
Because of Eqs.~(\ref{Ap11}) and~(\ref{Ap13}),  only two of the
three ${\bf R}_{\perp}$'s and two of the three $x$'s are
independent. We will choose as independent variables those for the
bosons, while the "nucleon" variables are expressed through them
as

\begin{equation}
\label{Ap14}
{\bf R}_{1\perp}=-({\bf R}_{2\perp}+{\bf R}_{3\perp}),\quad x_1=1-x_2-x_3.
\end{equation}

We also need to know the contractions of the four-momenta $p$ and
$k_n$ among themselves. Since for all $n$ we have $\omega\cd
R_n=0$, $R_n^2=-R_{n\perp}^2$, then, in full analogy with the
derivation of Eq.~(\ref{Ap4}),
\begin{equation}
\label{Ap15}
p\cd k_{n}=\frac{{R}_{n\perp}^2+M^2{x}_n^2+m_n^2}{2x_n},
\end{equation}
where $m_n$ is the mass related to the particle $n$. From the
condition $\omega\cd R_n=0$ follows
\begin{eqnarray}
\label{Ap16}
R_{n_1}\cd R_{n_2}&=&
-{\bf R}_{n_1\perp} {\bf R}_{n_2\perp}\nonumber \\
&=&(k_{n_1}-x_{n_1}p)\cd (k_{n_2}-x_{n_2}p),
\end{eqnarray}
which finally yields
\begin{eqnarray}
k_{n_1}\cd
k_{n_2}&=&\frac{x_{n_2}}{2x_{n_1}}({R}_{n_1\perp}^2+m_{n_1}^2)
\label{Ap17}\\
&&+\frac{x_{n_1}}{2x_{n_2}}({R}_{n_2\perp}^2+m_{n_2}^2)-{\bf
R}_{n_1\perp} {\bf R}_{n_2\perp}. \nonumber
\end{eqnarray}
It is also convenient to define the invariant energy squared of the three-body state:
\begin{equation}
\label{Ap18}
s_{123}\equiv \left (\sum_n k_n\right)^2=\sum_n \frac{{R}_{n\perp}^2+m_{n}^2}{x_n}.
\end{equation}
Omitting  the subscript 2 at ${\bf R}_{2\perp} $ and $x_2$, and
changing ${\bf R}_{3\perp}\to {\bf R}'_{\perp}$, $x_3\to x'$
everywhere in the above formulas, we transform the equations for
the vertex functions, obtained from the diagrams of
Fig.~\ref{scalarfig}, to those given by Eqs.~(\ref{eq11}).


\end{document}